\documentclass[aip,jcp,reprint,nobibnotes,superscriptaddress,floatfix, 
showpacs,citeautoscript,showkeys]{revtex4-1}

\usepackage{amsmath}

\DeclareMathAlphabet{\mathsf}{OT1}{phv}{b}{n}

\newcommand{\braket}[1]{\ensuremath{\left\langle#1\right\rangle}}

\newcommand{\crossVorg}{\ensuremath{%
         \setbox0=\hbox{$V$}
        V \kern-\wd0{\raise.3ex\hbox{$\relbar$}}}}

\newcommand{\crossVxx}[2]{%
	{\setbox0=\hbox{$#1#2V$}
         \setbox1=\hbox{$#1#2$}
         \setbox2=\hbox{$#1V$}
         \dimen1=\wd0
	 \advance\dimen1-\wd1
         \raise.2\ht0\hbox{$#1#2$}\kern-.4\wd0}}

\usepackage{pifont}

\usepackage{graphicx}
\usepackage[version=4]{mhchem}
\usepackage{textcomp}
\usepackage{amssymb}
\usepackage{amsbsy}
\usepackage{epstopdf}
\usepackage{amsmath}
\usepackage{bm}
\usepackage{longtable}
\usepackage{subfigure}
\usepackage{epsfig}
\usepackage{dcolumn}
\usepackage{latexsym}
\usepackage{textcomp}
\usepackage[Euler]{upgreek}
\usepackage{longtable}
\usepackage{multirow}

\usepackage{comment}

\newcommand{\be}{\begin{equation}}
\newcommand{\ee}{\end{equation}}

\makeatletter 
\gdef\@ptsize{0}
\let\@currsize\normalsize 
\makeatother 
\usepackage{setspace} 

\DeclareMathAlphabet\mathbfcal{OMS}{cmsy}{b}{n}
\DeclareMathAlphabet{\mathbfsf}{\encodingdefault}{\sfdefault}{bx}{sl}

\begin{document}
\title{Equilibrium binding energies from fluctuation 
theorems and force spectroscopy simulations}


\author{Emma Hodges}
\affiliation{Department of Chemical Engineering,
  Monash University, Melbourne, VIC 3800, Australia}
\affiliation{Monash Biomedicine Discovery Institute and Department of Microbiology, Monash University, VIC 3800, Australia} 

\author{B. M. Cooke}
\affiliation{Monash Biomedicine Discovery Institute and Department of Microbiology, Monash University, VIC 3800, Australia}

\author{E. M. Sevick}
\affiliation{Research School of Chemistry, Australian National
  University, Canberra ACT 0200, Australia}  

\author{Debra J. Searles} 
\affiliation{AIBN Centre
  for Theoretical and Computational Molecular Science, University of
  Queensland, Brisbane, QLD 4072, Australia}
\affiliation{School of
  Chemical and Molecular Biosciences, University of Queensland,
  Brisbane, QLD 4072, Australia} 
  
\author{B. D\"{u}nweg}
\affiliation{Department of Chemical Engineering, Monash University,
  Melbourne, VIC 3800, Australia} 
\affiliation{Max Planck Institute
  for Polymer Research, Ackermannweg 10, 55128 Mainz, Germany}
\affiliation{Condensed Matter Physics, TU Darmstadt, Hochschulstra{\ss}e 12, 
  64289 Darmstadt, Germany} 

\author{J.~Ravi~Prakash}
\email[Electronic mail: ]{ravi.jagadeeshan@monash.edu}
\affiliation{Department of Chemical Engineering, Monash University,
  Melbourne, VIC 3800, Australia}

\date{\today}

\begin{abstract}

Brownian dynamics simulations are used to study the detachment of a
particle from a substrate. Although the model is simple and generic,
we attempt to map its energy, length and time scales onto a specific
experimental system, namely a bead that is weakly bound to a cell and
then removed by an optical tweezer.  The external driving force arises
from the combined optical tweezer and substrate potentials, and
thermal fluctuations are taken into account by a Brownian force. The
Jarzynski equality and Crooks fluctuation theorem are applied to
obtain the equilibrium free energy difference between the final and
initial states. To this end, we sample non--equilibrium work
trajectories for various tweezer pulling rates. We argue that this
methodology should also be feasible experimentally for the envisioned
system. Furthermore, we outline how the measurement of a whole free
energy profile would allow the experimentalist to retrieve the unknown
substrate potential by means of a suitable deconvolution. The
influence of the pulling rate on the accuracy of the results is
investigated, and umbrella sampling is used to obtain the equilibrium
probability of particle escape for a variety of trap potentials.
\end{abstract}

\maketitle

\section{Introduction}
\label{sec:intro}

The adhesion of a cell to a substrate~\cite{kendall2010adhesion,
  bongrand2013studying,evans2007forces} occurs in a number of
biophysical contexts, and is hence a very important phenomenon to
study. Beyond its relevance for understanding biological phenomena in
general, many clinical applications in both diagnostics and
therapeutics fundamentally involve adhesion. Examples include: (i) the
sequestration of red blood cells in small blood vessels due to
infection with malaria~\cite{miller2002pathogenic,%
  turner1994immunohistochemical,rowe2009adhesion,suresh2005connections},
(ii) the growth of metastases in cancer~\cite{johnson1999cell,%
  hirohashi2003cell,thiery2003cell}, and (iii) the formation of
platelets at the site of a vascular
injury~\cite{hynes1992integrins}. A variety of experimental techniques
have been developed~\cite{neuman2008single} to measure the adhesive
properties of a single cell, such as atomic--force
microscopy~\cite{zhang2002force,benoit2000discrete,%
  zlatanova2000single,puech2006new}, surface--force apparatus
measurements~\cite{noy2007handbook}, micropipette
manipulation~\cite{chesla1998measuring,shao2004quantifying,Zhao01}, as
well as magnetic~\cite{dobson2008remote} and
optical~\cite{crick2014quantitation,grier2003revolution,%
  fallman2004optical} tweezers. All these methods subject the cell to
external time--dependent forces, with the aim of quantifying the
energetics of the binding.

The theoretical framework to analyse such experiments are the recently
developed non--equilibrium work theorems~\cite{evans2002fluctuation,%
  seifert2012stochastic}, most notably the Jarzynski
theorem~\cite{jarzynski1997nonequilibrium,Jarzynski:1997fu}, and
Crooks fluctuation theorem~\cite{Crooks:1998wk,%
  crooks1999entropy,Crooks:2000ke}, which have been used with great
success to interpret data from both computer simulations and
experiments~\cite{sandberg2015efficient,gapsys2015calculation,%
  carberry2004fluctuations,bustamante2005unfolding,gao2012non,wang2002}.
These theorems combine in a coherent fashion the three salient aspects
of the experiments, which are (i) the system's equilibrium statistical
physics (in particular the binding enthalpy), (ii) the fact that
time--dependent manipulation necessarily implies non--equilibrium
statistical physics (where the degree of deviation from equilibrium is
determined by the pulling speed or a similar parameter), and (iii) the
influence of thermal fluctuations. The central quantity of the
theorems is the non--equilibrium work that the external forces do on
the system. As soon as the external driving happens on a time scale
that is faster than the typical relaxation times of the system, the
non--equilibrium work is no longer simply given by the free energy
difference between final and initial state (as would be the case for
infinitely slow or quasi--static driving), but rather acquires a
dissipative contribution, which, as a result of thermal fluctuations,
has a statistical distribution of values. The theorems make detailed
statements on the relation between the probability distribution of the
non--equilibrium work and the underlying equilibrium free energies,
and are hence immensely useful to obtain the latter under experimental
conditions that cannot be considered as quasi--static. Essentially the
extraction of equilibrium properties from the non--equilibrium work
distribution is tantamount to reweighting the latter. Therefore, the
theorems, although in theory being applicable to a large class of
physical situations, have limitations in practice, since the
equilibrium free energy difference should not differ from the mean
non--equilibrium work by more than a few standard deviations --- and
this becomes more and more unfavourable both with increasing
dissipation and increasing system size. In practice, this means that a
reliable acquisition of equilibrium properties requires more and more
trajectories over which one needs to
average~\cite{jarzynski1997nonequilibrium,Jarzynski:1997fu}. In this
context, it should be noted that the theorems always consider
transitions from an equilibrium initial state to a final state, which
is typically out of equilibrium. These states are \emph{not} given by
some reaction coordinate of the system, but rather by the external
driving. Furthermore, we would like to mention that not only free
energies, but also other equilibrium properties (like e.~g. the
probability of attachment) can be obtained in an analogous fashion by
a suitable reweighting (or ``umbrella sampling'') procedure.

Binding between cells is complex and involves a slew of interactions,
which are both specific and non--specific~\cite{sackmann2006cell,%
  bell1984cell,gonnenwein2003generic}. The most important ingredient,
however, are bonds that arise from receptor--ligand pairs. Typically,
a single receptor--ligand interaction is fairly strong, i.~e., of
order of few $k_{\text{B}} T$ to $100 \, k_{\text{B}} T$, where
$k_{\text{B}}$ is Boltzmann's constant and $T$ the absolute
temperature at ambient
conditions~\cite{noy2007handbook,boal2012mechanics}, i.~e.
$k_{\text{B}}T \simeq 4 \text{pN\,nm}$. Moreover, cell adhesion will
in most cases involve many ligands, giving rise to a net total
interaction of typically several hundred $k_{\text{B}} T$.  The
mechanical detachment of a cell from a ``substrate'' to which it is
bound via receptor--ligand pairs (the latter can for example be
another cell, or a ligand--coated bead) is thus a very complex
process~\cite{sackmann2006cell,sackmann_physics_2014}. In a highly
simplified picture, we envision it to be roughly analogous to the
pulling--off of a plaster from skin, or to the pinch--off of a water
droplet from a dripping faucet. In optical tweezer
experiments~\cite{hodges2016inpreparation} we have observed that the
same external force can be sufficient to break some cell--substrate
pairs but insufficient to break others of the same type. In our
opinion, this provides an indication that the underlying dynamics
matter, and this will depend on details of variables such as the
number of receptor--ligand pairs, their density, and their geometrical
arrangement. At any rate, this means that a faithful modeling of
cell--substrate detachment or attachment would need to take into
account a large arrangement of receptor--ligand bonds, and their
(elastic) interactions. The single pair, in turn, is weak enough that
thermal fluctuations crucially contribute to its formation and
breaking.

As a first step in the modeling of micromechanical manipulation of
cell attachment and detachment, we focus in the present paper on the
case of just a single ligand--receptor pair. This is clearly the
easiest situation, since in principle this allows us to just consider
a single coordinate $x$ as a degree of freedom, which may be viewed as
the cell--substrate distance. This degree of freedom can then be
viewed as subject to (i) forces from the cell--substrate interaction,
(ii) forces from the time--dependent external pulling, and (iii)
thermal agitation. This situation is less artificial than one might
think at first glance, since it is experimentally possible to modify
the adhesive properties of cells through gene--knockout techniques
and/or inhibitors \cite{fried1998maternal,beeson2000adhesion,%
  ockenhouse1991molecular,glenister2002contribution,%
  cooke2002assignment}, such that receptor--ligand interactions are
systematically turned off. The aim of the present theoretical study is
to demonstrate that in this weak--binding situation the theorems can
actually be applied practically to obtain reliable results on free
energies, and, as a consequence, on the binding energetics. To do
this, we study the attachment or detachment process within the
framework of a very simple theoretical model, whose dynamics is
simulated by means of Brownian Dynamics. An important aspect here is
the fact that the simulation parameters (strength and range of
interactions, pulling speed) roughly match those of real
experiments. In the subsequent sections we will provide details on the
choice of parameters, and discuss the relation between the free
energies from the fluctuation theorems on the one hand, and the
binding forces on the other.

It should be emphasised that our numerical model is fairly generic and
therefore in principle applicable to any micromechanical manipulation
that detaches one object from another (or attaches it to it), as long
as this process can be described by a single reaction coordinate, and
involves energies that are roughly comparable with $k_{\text{B}}
T$. However, what we have principally in mind are experiments with
optical tweezers. We believe this technique has a great potential in
the future, since it is fairly non--invasive, and provides good
quantitative control over the external forces involved. For this
reason, we choose our parameters in rough accordance with a typical
tweezer experiment, and also use a nomenclature that refers to this
situation. More precisely, we think of a cell tightly ``glued'' to a
glass surface~\cite{crick2014quantitation}, while a ligand--coated
bead is moved due to the influence of a time--dependent (harmonic)
tweezer potential. The forces that the cell exerts on the bead are
then described by a fixed (not time--dependent) ``membrane
potential''.

It is worth noting that fluctuation theorems have already been used to
computationally calculate binding free energies in drug--receptor
systems~\cite{sandberg2015efficient,gapsys2015calculation}. These
computations involve \emph{deterministic} nonequilibrium molecular
dynamics of ligand--receptor pairs whose molecular properties, such as
Lennard--Jones parameters and force fields are known. In this paper,
the analysis of single cell detachment events will be described and
the usefulness of fluctuation theorems demonstrated, using data
generated by \emph{stochastic} simulation of a model cell and
substrate. Since the situation in the numerical study is fairly
similar to a typical experiment, we believe that this also
demonstrates the usefulness of the approach to experimentally estimate
the strength of binding --- with the caveat that the experiments will
be less accurate, since it is experimentally not possible to study
$\mathcal O(10^6)$ trajectories, as was done in the present
investigation.

The remainder of the paper is organised in the following manner:
First, details of the Langevin simulation will be presented, including
code validation. Second, the Jarzynski and Crooks fluctuation theorems
are shown to be valid for this two state system. As a result,
non--equilibrium work trajectories, calculated for the different trap
velocities, can be used to obtain the equilibrium free energy
difference between the final and the initial state. We will also
briefly outline (although this has not been done in the present work)
how this information can in principle be used to retrieve the membrane
potential, which in an experiment is of course unknown. Third,
limitations of numerical calculations using the fluctuation theorems
will be discussed and illustrated with the use of cumulants. Finally,
umbrella sampling will be used to derive equilibrium values such as
the probability of detachment or adhesion for a variety of different
trap potentials.
      
\section{Problem formulation}
\label{sec:Model}

\subsection{The model unbinding experiment}

\begin{figure}[th!]
\begin{center}
\includegraphics[width=8cm,height=!]{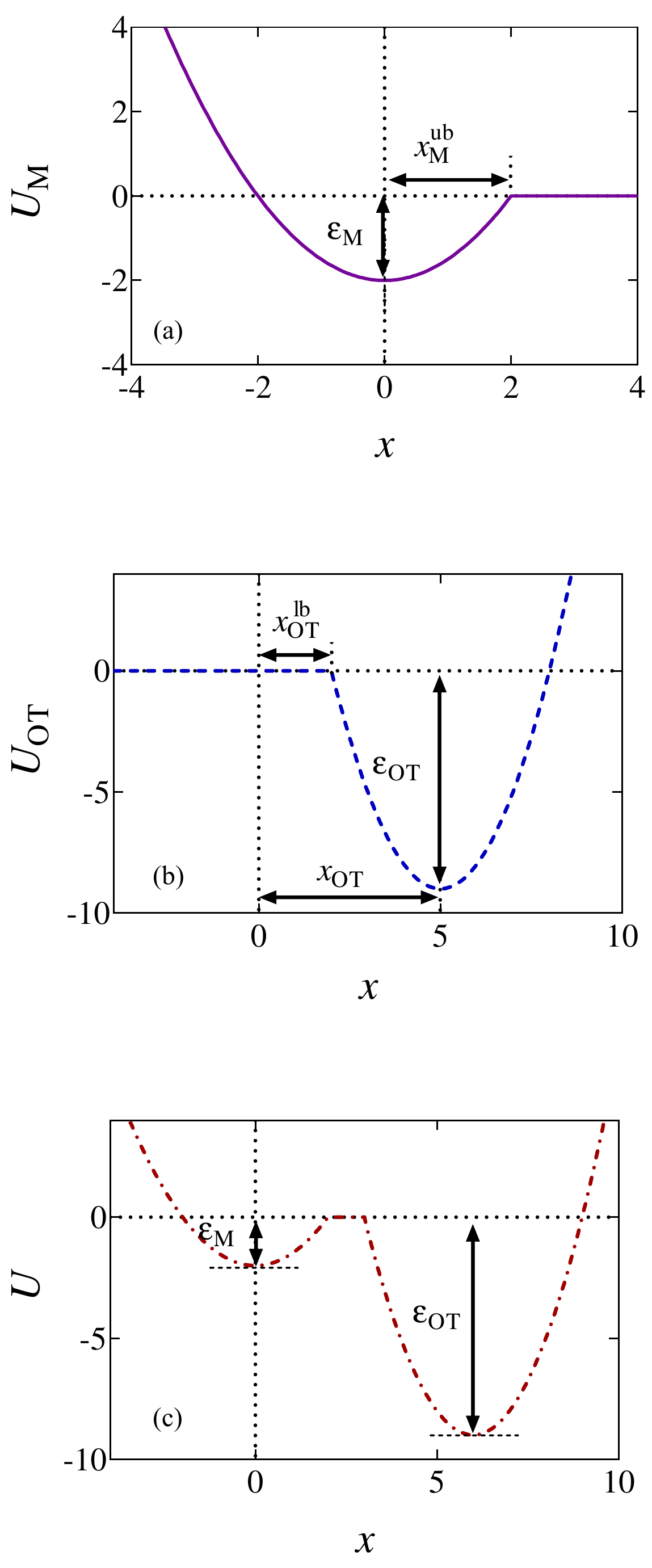} 
\end{center}
\vskip-20pt
\caption{(Color online) Schematic diagrams of the potentials. (a) The
  membrane potential (held stationary at all times). (b) The optical
  trap potential. The minimum, $x_{\text{OT}}$, changes linearly with
  time as the optical trap is moved at a constant speed
  $v_{\text{OT}}$ to a final position, $x_{\text{OT}}^{\text{final}} =
  6$. (c) The total potential, $U = U_{\text{M}} + U_{\text{OT}} $,
  experienced by the bead at some time $t>0$. In order to detach from
  the membrane the bead needs an energy greater than
  $\epsilon_{\text{M}}$, while in order for the bead to go from being
  unattached to attached, it would require an energy of order
  $\epsilon_{\text{OT}}$ or greater.}
\label{fig:pots}
\vskip-15pt
\end{figure}
Truncated harmonic potentials, as shown in Figs.~\ref{fig:pots}~(a)
and (b), are used to describe the interaction of the bead with both
the membrane and the optical trap. These potentials are made
dimensionless by scaling with the natural energy scale $k_{\text{B}}
T$, and defined by the expressions
\begin{equation}
\resizebox{0.36\textwidth}{!}{$
 U_{\text{M}}(x) = 
 \begin{cases} 
   \dfrac{1}{2} \, k_{\text{M}} \, x^2 - \epsilon_{\text{M}}
   & 
   \text{for} \quad x < x_{\text{M}}^{\text{ub}} \equiv 
   \sqrt{2 \epsilon_{\text{M}} / k_{\text{M}}}
   \\
   0 
   &
   \text{for} \quad x \geq  x_{\text{M}}^{\text{ub}}
   \end{cases}
$}   
   \label{eq:mempot}
\end{equation}
and
\begin{equation}
\resizebox{0.48\textwidth}{!}{$ U_{\text{OT}}(x) =
 \begin{cases} 
   0
   & 
   \text{for}  \quad  x < x_{\text{OT}}^{\text{lb}} \equiv x_{\text{OT}}
   - \sqrt{2 \epsilon_{\text{OT}} / k_{\text{OT}}} 
   \\
   \dfrac{1}{2} \, k_{\text{OT}} \, (x - x_{\text{OT}})^2 - \epsilon_{\text{OT}}
   &
   \text{for}  \quad  x \geq x_{\text{OT}}^{\text{lb}} ,
 \end{cases}
 $}
 \label{eq:otpot}
\end{equation}
where $U_{\text{M}}$ and $U_{\text{OT}}$ are the dimensionless
membrane and optical trap potential energies, respectively. The
distance $x$, measured from the fixed location of the minimum of the
membrane potential, is made dimensionless by scaling with a length
$\sqrt{k_{\text{B}} T / k_{\text{s}}}$, where $k_{\text{s}}$ is a
typical spring constant. We now choose the dimensionless parameters
$\epsilon_{\text{M}}$ and $\epsilon_{\text{OT}}$ of order unity, which
means that the involved energy scales are $\mathcal O(k_{\text{B}}
T)$, as in the envisioned experiments. Furthermore, we assume that the
spring constant $k_{\text{s}}$ is a value that corresponds to a
typical optical trap strength of $\mathcal O(10^{-3}
\text{pN/nm})$~\cite{noy2007handbook}, implying that $k_{\text{OT}}$ is
a dimensionless parameter of order unity. At ambient conditions,
$k_{\text{B}} T \simeq 4 \text{pN\,nm}$, meaning that a typical
thermal displacement within the trap (which is our unit of length) is
several tens of nanometers. The typical displacements that we observe
for cell detachment~\cite{hodges2016inpreparation} are of similar
order, and therefore we set $k_{\text{M}}$ as a parameter of order
unity as well.

The repulsive segment of the membrane potential ($-\infty < x \le 0$)
accounts for the impenetrability of the membrane to the bead, while
the attractive segment ($0 < x \le x_{\text{M}}^{\text{ub}}$)
represents the adhesive force exerted by the membrane on the bead
(Fig.~\ref{fig:pots}(a)). Beyond this distance, the bead detaches from
the membrane and the influence on the bead by the membrane potential
becomes negligible. Note that the minimum of the potential is held
fixed at the origin ($x=0$) for all time. Traditionally optical
tweezer potentials are represented by harmonic
wells~\cite{carberry2004,wang2002}. However, for investigations of
detachment or attachments one should take into account that the
optical trap has a finite range of attraction as well, such that a
truncated harmonic potential is more reasonable. In principle this
consideration holds for both branches $x < x_{\text{OT}}$ and $x >
x_{\text{OT}}$, where $x_{\text{OT}}$ is the (time--dependent)
location of the minimum of $U_{\text{OT}}$. However, it is crucially
important only for $x < x_{\text{OT}}$ because this controls the
energy barrier between the membrane and the trap potential. For $x >
x_{\text{OT}}$ we do \emph{not} truncate the tweezer potential, in
order to obtain finite expressions in the equilibrium statistical
mechanics of the system: If the total potential would exhibit an
infinite range of vanishing potential, then this region would
correspond to an infinite translational entropy, meaning that at any
finite temperature there could be no equilibrium adsorption of the
bead. Dynamically, this behavior would correspond to ``evaporation''
of the bead at sufficiently long times. It is therefore reasonable to
study the particle in a potential that results in a converging
partition function, and by this to strictly disregard such
``evaporation'' events (which, in a typical experiment, are anyway not
observed). These considerations lead us to assume a model tweezer
potential $U_{\text{OT}}$ as depicted in Fig.~\ref{fig:pots}~(b). The
total potential, $U(x) = U_{\text{M}}(x) + U_{\text{OT}}(x)$, at some
time $t>0$, is shown schematically in Fig.~\ref{fig:pots}~(c).

\begin{figure*}[th]
\centering
\includegraphics[width=18cm]{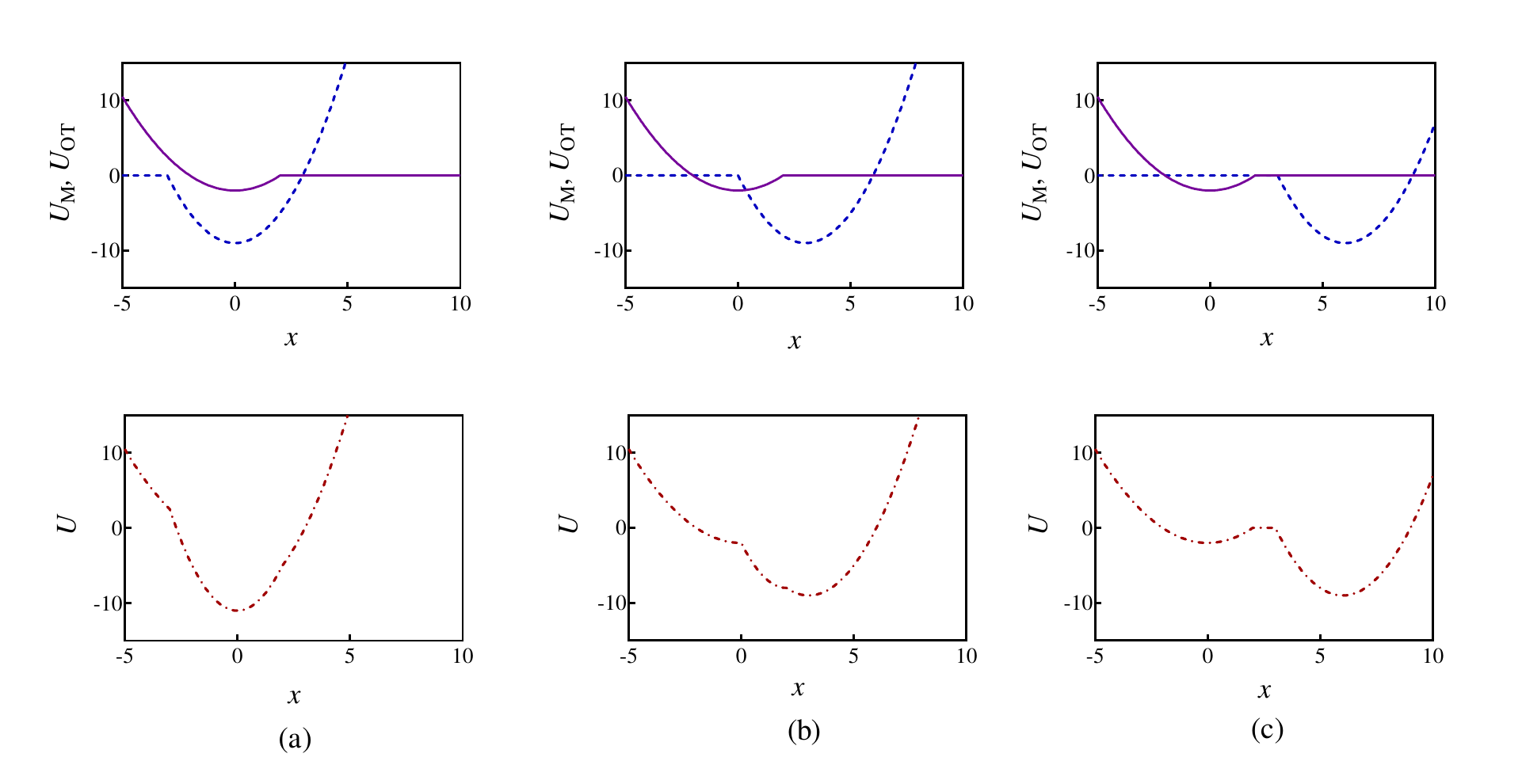}
\caption{Potential energy profiles when the optical trap minimum is at
  three different locations and corresponding to parameter set 1 in
  Table~\ref{table:par}. The first row shows the membrane (purple
  solid line) and optical trap (blue dashed line) potentials
  separately, whilst row two shows the summed potential (red
  dashed-dot line).  Potential shapes at: (a) $ x_{\text{OT}} = 0$,
  (b) $x_{\text{OT}} = 0.5 \, x_{\text{OT}}^{\text{final}}$, and (c) $
  x_{\text{OT}} = x_{\text{OT}}^{\text{final}}$. }
\label{fig:timepot}
\end{figure*}

The optical trap potential minimum is located at the origin at time $t
= 0$, i.~e., $x_{\text{OT}} (t=0) = 0$. At later times, the optical
trap is translated horizontally linearly with time, at varying speeds
$v_{\text{OT}}$ (i.~e., $x_{\text{OT}} (t) = v_{\text{OT}} \, t$), in
order to simulate the process of bead detachment by the optical
trap. The final position of the trap minimum is always at a fixed
location, $x_{\text{OT}}^{\text{final}} = 6$, regardless of the value
of $v_{\text{OT}}$. The summed potential $U$ is time dependent because
of the time dependence of the optical potential. For the purpose of
illustration, the shapes of the membrane and optical trap potentials,
along with the summed potential, during the course of the simulation,
at three different locations of the optical trap minimum are shown in
Fig.~\ref{fig:timepot}.
\begin{figure*}[th]
\begin{center}
\includegraphics[width=18cm,height=!]{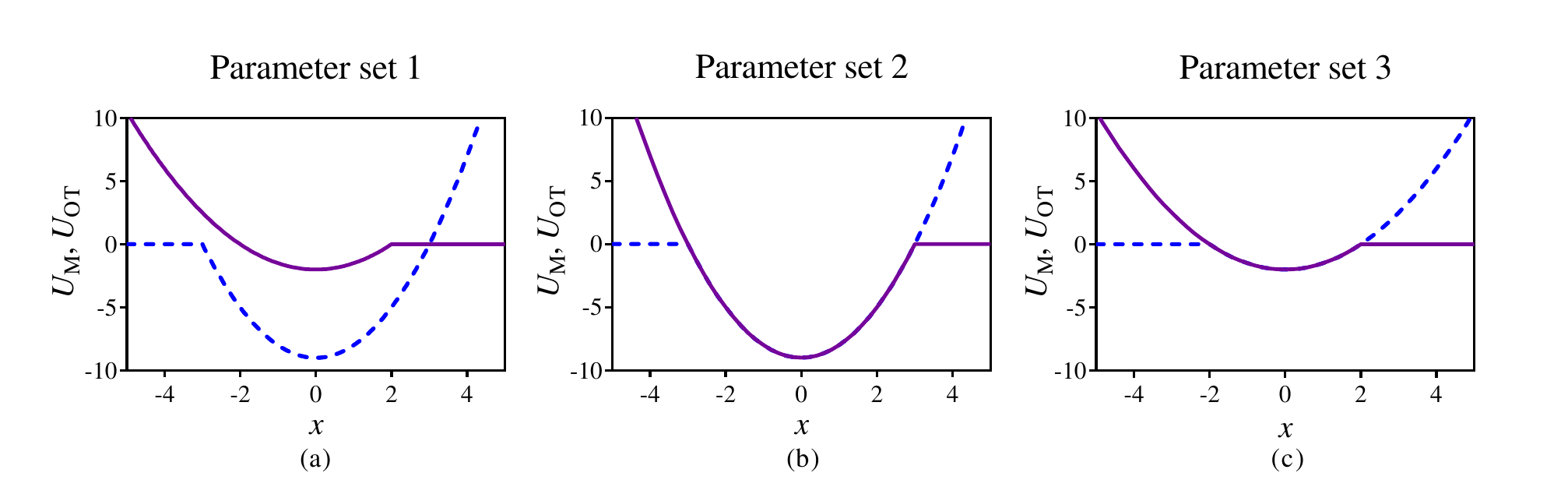} 
\end{center}
\vskip-20pt
\caption{(Color online) Snapshots of the membrane (purple solid line)
  and optical tweezer (blue dashed line) potentials at time $t=0$, at
  three different dimensionless values of well depths, and membrane
  and trap strengths, as given in Table~\ref{table:par}.}
\label{fig:par}
\end{figure*}

The relative ease of attachment and detachment is controlled by the
magnitudes of the barrier heights for the membrane
($\epsilon_{\text{M}}$) and the optical tweezer
($\epsilon_{\text{OT}}$) potentials, respectively, and also by their
respective strengths $k_{\text{M}}$ and $k_{\text{OT}}$. In order to
model different adhesive interactions between the bead and the
membrane, the barrier heights and spring constants can be changed
appropriately. In the present work, we choose three different sets of
values for these parameters (given in Table \ref{table:par}), allowing
different scenarios to be tested, as illustrated in
Fig.~\ref{fig:par}. In Fig.~\ref{fig:par}~(a), the membrane potential
is weaker than the optical trap in both strength and depth. In
Fig.~\ref{fig:par}~(b), both the potentials have the same strength and
depth, with the dimensional depth being of order 10 $k_{\text{B}} T$,
while in Fig.~\ref{fig:par}~(c), their dimensional depths are of order
$1 k_{\text{B}} T$. As will be seen subsequently, these three different
scenarios lead to considerably different adhesive behaviour.

\begin{table}[t]
\small
\begin{center}
\caption{ \label{table:par} Various dimensionless parameter values
  chosen to provide membrane and optical trap potentials with
  different depths and strengths}
  \vskip 10pt
\setlength{\tabcolsep}{10pt}
{\def\arraystretch{1.3}
 \begin{tabular*}{0.48\textwidth}{@{\extracolsep{\fill}}l l c c c }
 \hline
 \multirow{2}{*}{ }    &  \multicolumn{3}{c}{Parameter sets} 
 \\ 
 \cline{2-4}
 & $1$ &  $2$ & $3$  \\   \hline
 $k_{\text{M}}$  & $1$ & $2$ & $1$ \\ 
 $k_{\text{OT}}$& $2$ & $2$ & $1$ \\ 
 $\epsilon_{\text{M}}$ & $2$ & $9$ & $2$ \\ 
 $\epsilon_{\text{OT}}$ & $9$ & $9$ & $2$\\ 
 \hline
 \end{tabular*}
} 
\end{center}
\label{table:params}
\end{table}

\subsection{The Langevin equation}

In the absence of inertia, the time evolution of the particle's
position $x(t)$, subject to an external force due to the presence of
the membrane and optical potentials, and subject to thermal
fluctuations, is described by a Langevin equation
\begin{equation}
\label{eq:langevin}
\frac{d x}{d t} = F_{\text{ext}} + F_{\text{rand}}
\end{equation}
where the coordinate $x$ is dimensionless as described above, and time
is also made dimensionless by scaling with the typical time scale
$\zeta/k_{\text{s}}$, $\zeta$ being the friction coefficient of the
particle. $F_{\text{ext}}$ is the dimensionless external force due to
the combined potential, given by $F_{\text{ext}} = - \partial U /
\partial x$, while $F_{\text{rand}}$ is the dimensionless random force
(Gaussian white noise) with mean and variance
\begin{align}
 \label{eq:variance}
 \braket{F_{\text{rand}}} & = 0 \nonumber \\
 \braket{F_{\text{rand}}(t) F_{\text{rand}}(t')} & = 2 \, \delta (t-t')
\end{align} 
We use an Euler algorithm with a time step $\Delta t$,
\begin{equation}
  x (t + \Delta t) = x (t) + F_{\text{ext}} \, \Delta t 
  + \sqrt{2 \, \Delta t} \, r ,
\end{equation}
to numerically integrate the Langevin equation. Here $r$ is a random
number with $\braket{r} = 0$ and $\braket{r^2} = 1$. We use Gaussian
random numbers, applying the standard Box--Muller method.

Details of time step sizes and the number of trajectories used in the
simulations are given in the context of the various results discussed
below.

Assuming a typical bead radius of $4 \, \mu \text{m}$, and an aqueous
environment with viscosity $10^{-3} \text{Pa s}$, we find a Stokes
friction coefficient of $0.075 \times 10^{-3} \text{pN} \, \text{s} /
\text{nm}$, meaning that for a spring constant of $10^{-3} \text{pN} /
\text{nm}$ our unit of time is $0.075$ seconds.

The non--equilibrium aspect of the computer experiment comes in
through the finite pulling rate $v$ (the velocity at which the
location of the tweezer potential travels). For this we choose
dimensionless values between $0.01$ and $1$. In experimental units,
this means that even for the fastest process we pull the bead on a
time scale of not much less than roughly $0.1$ seconds, over a length
scale of a few ten nanometers, which means that the simulated process
corresponds well to experimentally feasible scales.

\subsection{Fluctuation theorems}
\label{fluct}
The initial and final states of our system are respectively defined as
(i) $x_{\text{OT}} = 0$, a situation where the tweezer potential keeps
the bead at a location close to the membrane, and (ii) $x_{\text{OT}}
= x_{\text{OT}}^{\text{final}}$ where it has moved the bead quite far
away from it, such that it feels only the force from the optical
trap. The fluctuation theorems are concerned with the free energy
difference $\Delta F$ between these two states.

If the unbinding is carried out isothermally and infinitesimally
slowly, then $\Delta F$ is equal to the work $W$ performed during the
process. On the other hand, if the unbinding experiment is carried out
at a finite rate over a period of time $t_{\text{D}}$, the work
performed will not be unique. Rather, an ensemble of such unbinding
experiments will lead to a distribution of work values, $P_{\text{F}}
(W)$ (where the subscript `F' indicates the experiment is carried out
in the \textit{forward} direction, from the cell and bead being bound
together to being unbound). Note that in this scenario, it is possible
that at the end of the experiment, the bead remains close to the cell,
even though work has been performed. In the quasi--static limit
$t_{\text{D}} \to \infty$, $P_{\text{F}} (W) \to \delta (W - \Delta
F)$. For finite rates of detachment, however,
\begin{equation}
\label{workineq}
 \braket{W} = \int dW \,  W \, P_{\text{F}} (W) \geq \Delta F .
\end{equation}
The great advance that has been made with the recently developed
fluctuation theorems is that, contrary to the suggestion of
Eqn.~(\ref{workineq}), a knowledge of the non--equilibrium work
distribution is sufficient to determine the equilibrium free energy
$\Delta F$ exactly.

The two fluctuation theorems that are primarily used in this work are
the Crooks fluctuation
theorem~\cite{Crooks:1998wk,crooks1999entropy,Crooks:2000ke}, and the
Jarzynski
equality~\cite{jarzynski1997nonequilibrium,Jarzynski:1997fu}. Both
these theorems are based on the following set of assumptions. The
system, whose dynamics is in our case stochastic and Markovian, is
driven by an external perturbation from an initial equilibrium state,
to a final state that is not necessarily at equilibrium. The external
parameter driving the perturbation at a finite rate from the initial
to the final state is denoted by $\lambda$, with values $\lambda_0$ in
the initial equilibrium state, and $\lambda_{\text{f}}$ in the final
state.

The Crooks fluctuation theorem states
that~\cite{Crooks:1998wk,crooks1999entropy,Crooks:2000ke}
\begin{equation}
\frac{P_\text{F}(W)}{P_\text{R}(-W)} = \exp{\left[ W-\Delta F \right] } ,
\label{eq:FTcft}
\end{equation}
where both the work and the free energy have been made dimensionless
by scaling with our energy unit $k_{\text{B}} T$. The distribution
$P_\text{F}(W)$ is the probability that the work of magnitude $W$ is
performed in perturbing the system from an initial equilibrium state
with $\lambda=\lambda_0$ to a final state with
$\lambda=\lambda_\text{f}$ in a finite time $t_\text{D}$, while
$P_\text{R}(-W)$ is the probability that work of the same magnitude
but opposite sign will be performed on perturbing the system in the
reverse path, from an equilibrium state with
$\lambda=\lambda_\text{f}$ to a state with $\lambda=\lambda_0$, over
the same length of time.

Equation \ref{eq:FTcft} clearly suggests that the value of work $W^*$
at which $P_\text{F}(W^*)=P_\text{R}(-W^*)$, is nothing but the
equilibrium free energy difference between the initial and final
states. We use this result subsequently in order to estimate the free
energy of binding.

The Jarzynski equality in its original
form~\cite{jarzynski1997nonequilibrium,Jarzynski:1997fu} only
considers perturbations from $\lambda_0$ to $\lambda_\text{f}$, and
states that
\begin{equation}
 \braket{e^{- W}}_\text{F} = e^{- \Delta F} ,
 \label{eq:FTje}
\end{equation}
where the subscript `F' on the ensemble average on the left hand side
indicates an average over forward trajectories. While the ensemble
average of the non--equilibrium work is always greater than the
equilibrium free energy for finite rates of system perturbation,
Jarzynski's equality states that an ensemble average of the
exponential of $(-W)$ can be used to directly evaluate the equilibrium
free energy. As will be seen subsequently, however, driving the system
from $\lambda_0$ to $\lambda_\text{f}$ at increasingly rapid rates
leads to a widening of the distribution $P_\text{F}$, and consequently
requires larger and larger ensembles to obtain an accurate estimate of
$\Delta F$.
\begin{widetext}
The experimental and practical relevance of these relations becomes
clear when considering the defining relation for the free energy,
\begin{equation}
\exp ( - \Delta F ) = \frac{
\int_{-\infty}^{+\infty} dx \, \exp (- U_{\text{M}} (x) ) 
\exp (- U_{\text{OT}} (x - x_{\text{OT}}^{\text{final}}) ) 
}{
\int_{-\infty}^{+\infty} dx \, \exp (- U_{\text{M}} (x) ) 
\exp (- U_{\text{OT}} (x - 0) ) 
} 
,
\end{equation}
where we emphasise that the tweezer potential depends on the
difference $x - x_{\text{OT}}$. Now, the fluctuation theorems permit
us to determine the free energy not only for the final state of the
tweezer potential, but also for any intermediate state
$x_{\text{OT}}^{\text{interm}}$. We thus find
\begin{equation}
\exp ( - \Delta F (x_{\text{OT}}^{\text{interm}}) ) = \frac{
  \int_{-\infty}^{+\infty} dx \, \exp (- U_{\text{M}} (x) ) \exp (-
  U_{\text{OT}} (x - x_{\text{OT}}^{\text{interm}}) ) }{
  \int_{-\infty}^{+\infty} dx \, \exp (- U_{\text{M}} (x) ) \exp (-
  U_{\text{OT}} (x) ) } 
 .
\end{equation}
Defining
\begin{equation}
  \phi(x - x_{\text{OT}}) = \exp (- U_{\text{OT}} (x - x_{\text{OT}}) ) ,
\end{equation}
which we can assume to be known since the properties of the
optical trap are known, and
\begin{equation}
  \psi (x) = \exp (- U_{\text{M}} (x) )
  \left[ 
\int_{-\infty}^{+\infty} dx \, \exp (- U_{\text{M}} (x) ) 
\exp (- U_{\text{OT}} (x) ) \right]^{-1} 
,
\end{equation}
\end{widetext}
which is not known, we can write
\begin{equation}
  \exp ( - \Delta F (x_{\text{OT}} ) ) = \int_{-\infty}^{+\infty} dx \,
  \phi(x - x_{\text{OT}}) \psi(x) .
\label{convol}  
\end{equation}

In other words, the exponential of the free energy profile, which is
experimentally accessible via the fluctuation theorems, is nothing but
the convolution of the known Boltzmann factor of the tweezer potential
with the unknown Boltzmann factor of the membrane
potential. Therefore, it should be possible to retrieve the latter by
just a numerical deconvolution, assuming that the free energy profile
is known with sufficient accuracy. More precisely, the procedure
yields $U_{\text{M}}$ up to an unknown constant, which is however
obviously irrelevant. Mapping out the membrane potential is, in our
opinion, the ideal goal of such experiments. In the present work, we
do not perform this program, but rather confine ourselves to the
simpler task of just determining $\Delta F$ for a single final state.

\subsection{Non--equilibrium work}

The application of the fluctuation theorems requires the determination
of the distribution of work $P_{\text{F}} (W)$ when the system is
driven from $\lambda_0$ to $\lambda_{\text{f}}$ in the forward path,
and the distribution $P_{\text{R}} (W)$ when the path is
reversed. Following the arguments of
Jarzynski~\cite{Jarzynski:1997fu}, we introduce the function
$H_\lambda (x)$, as the energy of the system for any fixed value of
$\lambda$, where $x(t)$ is the stochastic phase-space trajectory that
describes the time evolution of the system, which depends on the time
dependence of the external parameter $\lambda$. The total work
performed on the system, when it evolves from $\lambda = \lambda_0$ to
$\lambda = \lambda_\text{f}$, in a time period $t_{\text{D}}$,
is~\cite{Jarzynski:1997fu}
\begin{equation}
\label{eq:workdef}
W = \int_0^{t_\text{D}} \! dt^\prime \, {\dot \lambda} 
\frac{\partial H_\lambda}{\partial \lambda} \left( x(t^\prime) \right) 
\end{equation}
where ${\dot \lambda} = d \lambda/dt$. The stochastic phase--space
trajectory $x(t)$ of the bead is determined here by solving the Langevin
equation~(\ref{eq:langevin}). In the model system considered here, the only component of
the system's energy that depends on the external driving parameter
$\lambda \, (= x_\text{OT})$, is the potential energy of the trap,
$U_\text{OT}$. As a result, ${\partial H_\lambda}/{\partial \lambda} =
{\partial U_\text{OT}}/{\partial x_\text{OT}}$, and ${\dot \lambda} =
d x_\text{OT}(t)/dt = v_\text{OT}$. From Eqn.~(\ref{eq:otpot}), for $
x \geq x_\text{OT}^\text{lb}$, since
\begin{equation}
\label{eq:Fot}
F_{\text{OT}} (x) = - \frac{\partial U_{\text{OT}}}{\partial x}
= \frac{\partial U_{\text{OT}}}{\partial x_{\text{OT}}} 
= - k_{\text{OT}} \, (x - x_{\text{OT}} ) 
\end{equation}
it follows that
\begin{equation}
\label{eq:work}
W = \int_0^{t_{\text{D}}} \!\! dt^\prime \, v_{\text{OT}} \, 
F_{\text{OT}} \left( x(t^\prime) \right) . 
\end{equation}
Equations~(\ref{eq:Fot}) and~(\ref{eq:work}) are used here to
calculate the work done on the bead when the optical trap is
translated from $x_\text{OT} = 0$ to $ x_{\text{OT}} =
x_{\text{OT}}^{\text{final}}$, at all times $t$ at which the bead's
location satisfies $x(t) \geq x_{\text{OT}}^{\text{lb}}$. At other
times, when the force of the optical trap on the bead is zero, the
contribution to the work is zero. At any time $t$ during the course of the Langevin simulation, the accumulated work until time $t$ is calculated by numerically evaluating  the integral in Eqn.~(\ref{eq:work}) from $t^\prime=0$ to $t^\prime=t$. Since the typical time steps used in the simulation are very small ($\Delta t = 10^{-4}$ to $\Delta t = 10^{-3}$), a simple rectangular method was used to carry out the quadrature, where at each time step, the accumulated work at the end of the previous time step is augmented by the product of the value of the integrand at the beginning of the time step with $\Delta t$.

\subsection{Analytical evaluation of the free energy}
\label{analfe}

For the simple model considered here, the free energy difference
between the initial and final states can be evaluated analytically
exactly, and is given by
\begin{equation}
\label{eq:df}
\Delta F_\text{anal} = F_{\lambda_\text{f}} - F_{\lambda_0} 
= - \ln \frac{Z(x_\text{OT} = x_\text{OT}^\text{final})}{Z(x_\text{OT} = 0)} ,
\end{equation}
where the respective partition functions are given by the expressions
\begin{multline}
Z(x_\text{OT} = 0) 
= \int_{-\infty}^{x_\text{OT}^\text{lb}}  \! dx \exp{\left[-\left(\dfrac{1}{2} 
\, k_{\text{M}} \, x^2 - \epsilon_\text{M}\right)\right]}\\
+ \int_{x_\text{OT}^\text{lb}}^{x_\text{M}^\text{ub}}  \! dx \exp{\left[-\left(\dfrac{1}{2} 
\, k_{\text{M}} \, x^2 - \epsilon_\text{M}+\dfrac{1}{2} \, k_{\text{OT}} \, x^2 
- \epsilon_\text{OT} \right)\right]} \\
+ \int_{x_\text{M}^\text{ub}}^{\infty}  \! dx \exp{\left[-\left(\dfrac{1}{2}
\, k_{\text{OT}} \, x^2 - \epsilon_\text{OT} \right)\right]} ,
\end{multline}
\begin{multline}
Z(x_\text{OT} = x_\text{OT}^\text{final}) 
= \int_{-\infty}^{x_\text{M}^\text{ub}}  \! dx \exp{\left[-\left(\dfrac{1}{2} 
\, k_{\text{M}} \, x^2 - \epsilon_\text{M}\right)\right]} \\
+ (x_{\text{OT}}^{\text{lb}} - x_{\text{M}}^{\text{ub}})
\\
+ \int_{x_\text{OT}^\text{lb}}^{\infty}  \! dx \exp{\left[-\left(\dfrac{1}{2}
\, k_{\text{OT}} \, (x-x_\text{OT}^\text{final})^2 - \epsilon_\text{OT} \right)\right]}.
\end{multline}
The bounds on the integrals in the expressions above can be understood
from the schematic representations of the potentials in
Figs.~\ref{fig:pots} and~\ref{fig:timepot}.

These integrals can be evaluated analytically, and give rise to the
following expressions for the partition functions of the initial and
final states, respectively,
\begin{multline}
\label{anal0}
Z(x_\text{OT} = 0)
= \frac{\sqrt{\pi /2}}{\sqrt{k_{\text{M}}}} \exp{(\epsilon_{\text{M}})} 
\left[\text{erf}\left(\frac{x_{\text{OT}}^{\text{lb}}\sqrt{k_{\text{M}}}}{\sqrt{2}}\right)+1\right]\\
+ \frac{\sqrt{\pi /2}}{\sqrt{k_{\text{M}}+k_{\text{OT}}}} 
        \exp{(\epsilon_{\text{M}}+\epsilon_{\text{OT}})}
\left[\text{erf}\left(\frac{x_{\text{M}}^{\text{ub}}\sqrt{k_{\text{M}}+k_{\text{OT}}}}{\sqrt{2}}\right) \right. \\
  \left.   -\text{erf}\left(\frac{x_{\text{OT}}^{\text{lb}}\sqrt{k_{\text{M}}+k_{\text{OT}}}}{\sqrt{2}}\right)
+1\right]\\
+ \frac{\sqrt{\pi /2}}{\sqrt{k_{\text{OT}}}} \exp{(\epsilon_{\text{OT}} )}
\left[\text{erfc}\left(\frac{x_{\text{M}}^{\text{ub}}\sqrt{k_{\text{OT}}}}{\sqrt{2}}\right)\right] 
\end{multline}
\begin{multline}
\label{anal1}
Z(x_{\text{OT}} = x_{\text{OT}}^{\text{final}}) 
=\frac{\sqrt{\pi /2}}{\sqrt{k_{\text{M}}}} \exp{(\epsilon_{\text{M}})}
\left[\text{erf}
\left(\frac{x_{\text{M}}^{\text{ub}} \sqrt{k_{\text{M}}}}{\sqrt{2}}\right) 
+ 1 \right] \\
+ \left[ x_{\text{OT}}^{\text{lb}} - x_{\text{M}}^{\text{ub}} \right] \\
+ \frac{\sqrt{\pi /2}}{\sqrt{k_{\text{OT}}}} \exp{(\epsilon_{\text{OT}})}
\left[\text{erfc}
\left(\frac{( x_{\text{OT}}^{\text{lb}} - x_{\text{OT}}^{\text{final}} )
\sqrt{k_{\text{OT}}}}{\sqrt{2}}\right)
\right]
\end{multline}
Equations~(\ref{anal0}) and~(\ref{anal1}) can be used along with
Eqn.~(\ref{eq:df}) to obtain the exact value of the free energy
difference between the initial and final state for any choice of
parameter values in the potentials $U_{\text{M}}(x)$ and
$U_{\text{OT}}(x)$. Free energy differences for the particular choice
of values listed in Table~\ref{table:par} as parameter sets 1, 2 and
3, are given in Table~\ref{table:deltaf}. They are used to evaluate
the accuracy of the free energy differences predicted by the Crooks
and Jarzynski fluctuation theorems.

\section{Results and Discussion}
 \label{sec:results}
 
\subsection{ Code validation}
\label{A:valid}

\begin{figure*}[t]
\begin{center}
\begin{tabular}{cc}
\resizebox{9.25cm}{!} {\includegraphics*[width=9.5cm]{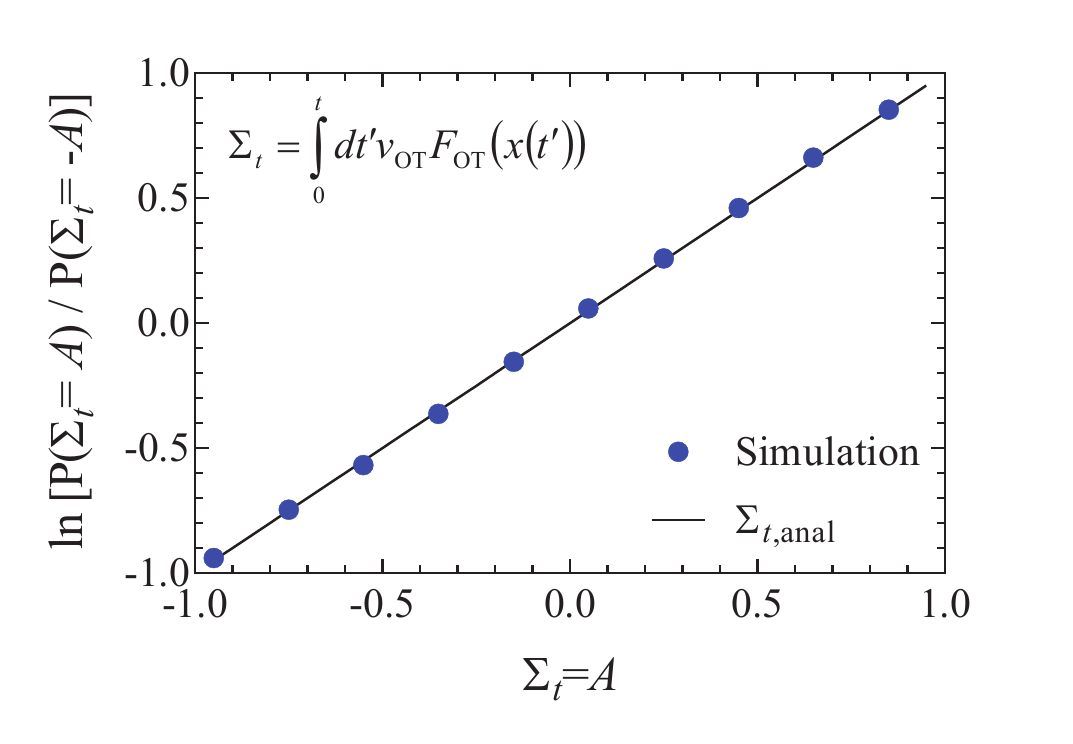}} & 
\resizebox{9.25cm}{!} {\includegraphics*[width=9.5cm]{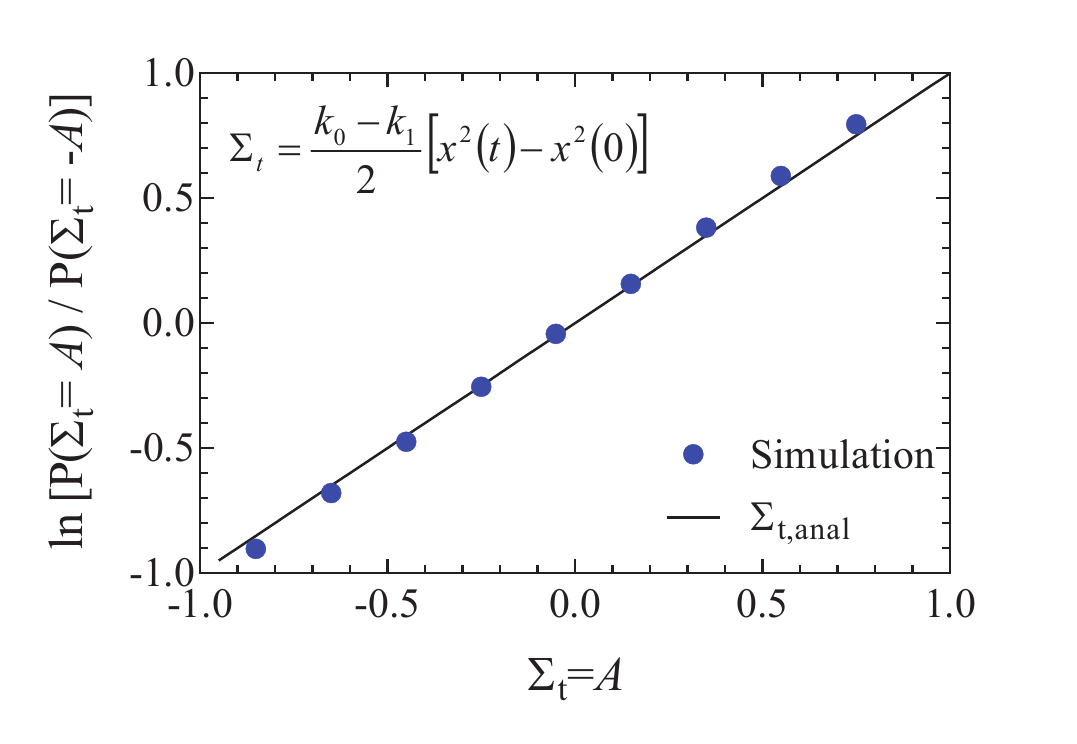}}\\
(a) & (b) 
\end{tabular}
\end{center}
\vskip-15pt
\caption{\label{fig:tft} Validation of code through demonstration of
  the Evans-Searles transient fluctuation theorem. Natural log of the
  number ratio of trajectories with entropy production $\Sigma_{t}$ to
  those with entropy production $-\Sigma_{t}$ versus $\Sigma_t$
  (filled circles), found from $2 \times 10^6$ trajectories. Lines are
  drawn with slope of 1 as predicted by the TFT (indicated as
  $\Sigma_{t, \text{anal}}$ in the figure legend). (a) Study 1
  (\citet{wang2002}). A line of best fit through simulation data has a
  slope $1.007 \pm 0.004$. (b) Study 2 (\citet{carberry2004}). A line
  of best fit through simulation data has a slope $1.058 \pm 0.002$. }
\end{figure*}
\begin{figure*}[!ht]
\begin{center}
\begin{tabular}{cc}
\resizebox{9.25cm}{!} {\includegraphics*[width=9.5cm]{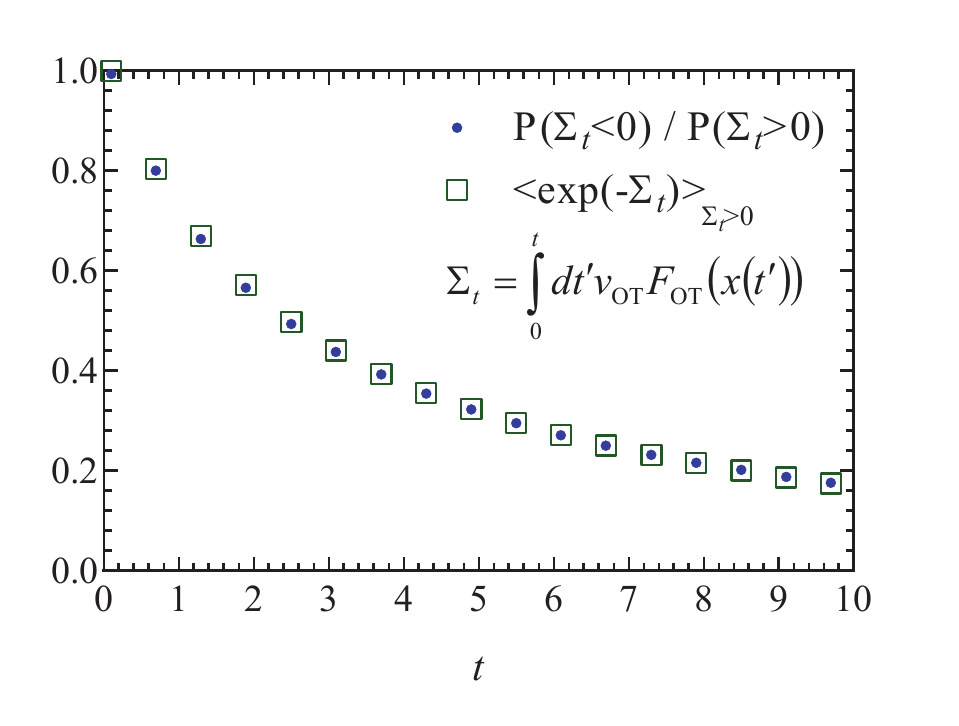}} & 
\resizebox{9.25cm}{!} {\includegraphics*[width=9.5cm]{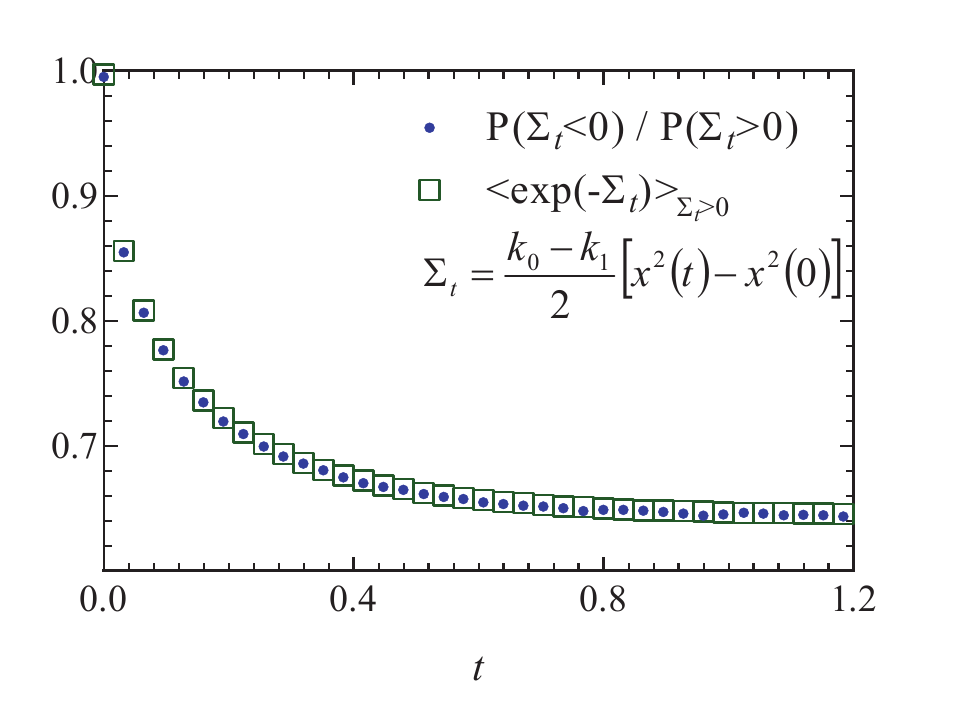}}\\
(a) & (b)  \\
\end{tabular}
\end{center}
\vskip-15pt
\caption{\label{fig:itft} Validation of code through demonstration of
  the Evans-Searles integrated fluctuation theorem. The number ratio
  of entropy consuming ($\Sigma_{t} < 0$) trajectories to entropy
  producing ($\Sigma_{t} > 0$) trajectories (filled circles), and the
  entropy production averaged over entropy producing trajectories,
  $\braket{\exp (-\Sigma_t)}_{\Sigma_t>0}$ (empty squares), versus
  time, found from $2 \times 10^6$ trajectories. (a) Study 1
  (\citet{wang2002}). (b) Study 2 (\citet{carberry2004}).}
\end{figure*}

In order to validate the predictions of the current algorithm,
comparisons were carried out with the results of two earlier studies
which demonstrated the Evans-Searles fluctuation theorems using
experiments and simulations involving an optical
trap~\cite{wang2002,carberry2004}. The \emph{transient fluctuation
  theorem} (TFT) of Evans and
Searles~\cite{Evans:1994go,Evans:2002gy,evans2002fluctuation} states
that
\begin{equation}
\frac{P(\Sigma_t=A)}{P(\Sigma_t=-A)}=\exp(A) ,
\label{eq:tft}
\end{equation}
while the \emph{integrated form} of the transient fluctuation theorem
(ITFT) states that
\begin{equation}
  \frac{P(\Sigma_t<0)}{P(\Sigma_t>0)}=
  \braket{\exp (-\Sigma_t)}_{\Sigma_t>0} .
\label{eq:itft}
\end{equation}
Here, $\Sigma_t$ is the dissipation function, which is a dimensionless
measure of the total entropy production that occurs along the system's
trajectory, over time $t$. It assumes different forms depending on the
system under consideration. The TFT relates the probability of
observing a trajectory with entropy \emph{production}, $\Sigma_t = A$,
to the probability of observing a trajectory with the
\emph{consumption} of the same magnitude of entropy, $\Sigma_t = -
A$. On the other hand, the integrated version of the theorem specifies
a relationship between the frequency of entropy-consuming trajectories
to that of entropy-producing trajectories, with the average on the
right hand side of Eqn.~(\ref{eq:itft}) carried out over only
entropy-producing trajectories.

In the first study considered here,\ ~\citet{wang2002} examined the
trajectory of a colloidal particle captured in an optical trap
translated at a uniform velocity relative to the surrounding
medium. They experimentally demonstrated the validity of the ITFT, and
also carried out molecular dynamics simulations to show that the
predictions of both the TFT and the ITFT were correct. In the second
study,\ ~\citet{carberry2004} observed the time-dependent relaxation
of a colloidal particle subjected to a step change in the strength of
a stationary optical trap. In this case, they were able to
experimentally demonstrate the validity of both the TFT and the ITFT.

We have carried out Langevin simulations of these two previously
studied applications of the Evans-Searles fluctuation theorems in
order to ensure that our algorithm was implemented correctly. In both
these examples, only a single optical trap is involved.  As a
consequence, the external force (in Eqn.~(\ref{eq:langevin})) on the
colloidal particle due to the optical trap is given by,
\begin{equation}
F_\text{ext} (t) = - k_\text{OT} \, (x(t) - x_\text{OT} (t))
\label{fext}
\end{equation}
where $k_\text{OT}$ and $x_\text{OT} (t)$ assume different expressions
in the two studies. As mentioned earlier, the dissipation function
$\Sigma_t$ is also different in the two cases. The relevant
expressions are listed below.

\noindent \textbf{Study 1} (\citet{wang2002}):
\begin{align}
k_\text{OT}&  = \text{constant} \nonumber \\
x_\text{OT} (t)&  = x_\text{OT} (0) + v_\text{OT} \, t \nonumber \\
\Sigma_t&  = \int_0^{t} \!\! dt^\prime \, v_\text{OT} \, 
F_\text{OT}\left( x(t^\prime) \right) \nonumber
\end{align}
where $F_\text{OT} (x)$ is given by Eqn.~(\ref{eq:Fot}).

\noindent \textbf{Study 2} (\citet{carberry2004}):
\begin{align}
k_\text{OT}&  =  k_0 + (k_1 - k_0) H (t) \nonumber \\
x_\text{OT} (t)&  = \text{constant} = 0 , \, \, \text{for all} \, \, t \nonumber \\
\Sigma_t&  = \frac{k_0 - k_1}{2} \, \left[ x^2(t) - x^2 (0) \right] \nonumber
\end{align}
where $H(t)$ is the Heaviside step function, and $k_0$ and $k_1$ are
constants equal to the optical trap strength before and after the step
change, respectively.

\begin{figure*}[t]
\begin{center}
\includegraphics[width=18cm]{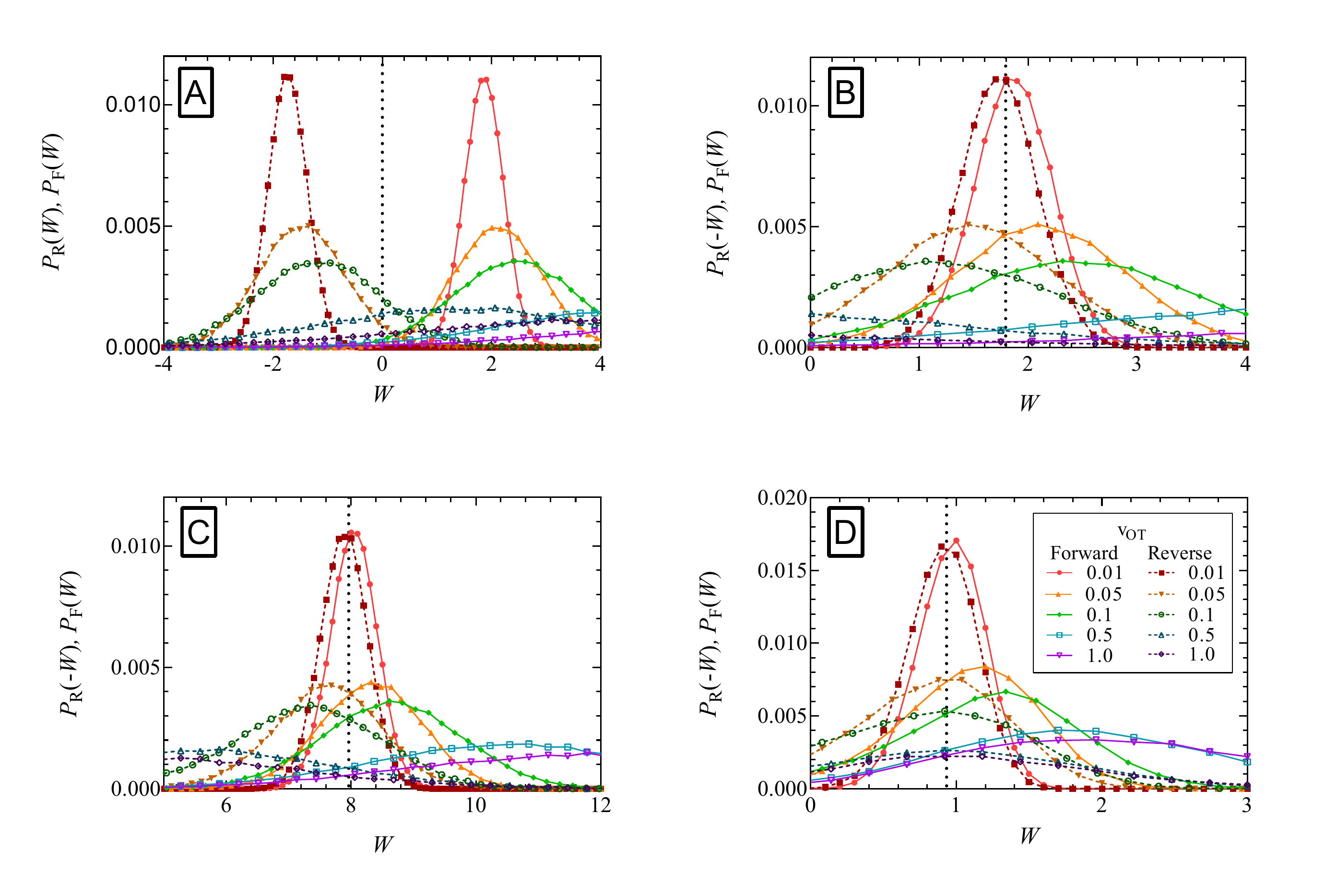}
\end{center}
\vskip-20pt
\caption{Evaluation of the equilibrium free energy using the
  Crooks fluctuation theorem for the three sets of potential parameter
  values listed in Table \ref{table:par}. In panel A, the probability
  of work $W$ being performed in the \emph{forward} path
  ($P_{\text{F}}(W)$) is plotted alongside the distribution of work
  values in the \emph{reverse} path ($P_{\text{R}}(W)$) for parameter
  set 1, for the trap velocities $v_{\text{OT}} = \{0.01, 0.05, 0.1,
  0.5, 1 \}$. In panels B (parameter set 1), C (parameter set 2), and
  D (parameter set 3), $P_{\text{F}}(W)$ is plotted alongside
  $P_{\text{R}}(- W)$. Note that the equilibrium free energy $\Delta F =
  W^*$, where $W^*$ is the value of work at which $P_{\text{F}}(W^*) =
  P_{\text{R}}(- W^*)$ (indicated by the dotted vertical lines in panels
  B--D).}
\label{fig:crooks}
\end{figure*}

The Langevin simulation of both these cases was carried out with $2
\times 10^{6}$ trajectories, using a time step of $10^{-4}$. In both
cases, after an initial equilibration time of $10^{4}$ time steps, the
distribution of particle positions was checked to see if the
respective equilibrium distribution functions were obeyed. In Study 1,
after equilibration, the optical trap was translated with a constant
velocity $v_\text{OT} =0.5$, from time $t=0$ to $t=10$, with a
constant trap strength $k_\text{OT} = 1$. In Study 2, after
equilibration, the optical trap strength was changed discontinuously
from $k_0 = 1$ to $k_1 = 2$ at time $t=0$, and the simulation
continued until $t=10$. The position of the colloidal particle at time
$t=0$ is taken to be $x(0)$. Figures~\ref{fig:tft} and~\ref{fig:itft}
summarise the results of the validation studies.

In order to demonstrate the TFT a histogram of the values of the
dissipation function $\Sigma_t$ at the end of the simulation was
constructed over the $2 \times 10^{6}$ trajectories. If $N_i$ is the
number of trajectories with dissipation function between $\Sigma_{t,i}
\pm \Delta/2$ (where $\Delta =0.1$ is the size of the histogram bin,
and $\Sigma_{t,i} = i \, \Delta$), then the ratio of probabilities on
the left hand side of Eqn.~(\ref{eq:tft}) can be evaluated from
$\left(N_i/N_{-i} \right)$. Figures~\ref{fig:tft}(a) and
\ref{fig:tft}(b) show the natural log of the ratio of the
probabilities obtained in this manner for both the studies, plotted
against the value of $\Sigma_{t}$. Also shown in the figures is a line
of slope unity, which represents the prediction of the TFT.

The ITFT is demonstrated for the two studies in
Figs.~\ref{fig:itft}(a) and \ref{fig:itft}(b), respectively, by
plotting the ratio of the number of entropy consuming trajectories
($\Sigma_{t} < 0$) to the number of entropy producing ($\Sigma_{t} >
0$) trajectories as a function of time, along with the time dependence
of the entropy production averaged over the subset of $2 \times
10^{6}$ trajectories in which entropy is produced.

\subsection{Crooks fluctuation theorem}

Simulations were carried out with the three sets of parameter values
listed in Table \ref{table:par} for the membrane and optical trap
potentials, with a time step size $\Delta t = 10^{-3}$. Rather than
running the simulations for an initial equilibration period, the
positions of the bead at time $t=0$ were chosen such that they
satisfied the known initial equilibrium distribution functions. Two
kinds of simulations were carried out. The first kind, that generated
\emph{forward} trajectories, started at time $t=0$ with the optical
trap minimum at $x_{\text{OT}} = 0$, followed by the trap minimum being
translated with a uniform velocity $v_{\text{OT}}$ until it was located
at $x^{\text{final}}_{\text{OT}}$ at time $t=t_{\text{D}}$. The set of
optical trap velocities $v_{\text{OT}} = \{0.01, 0.05, 0.1, 0.5, 1 \}$
was used. Note that $t_{\text{D}}$ depends on the value of $v_\text{OT}$
since the location $x^{\text{final}}_{\text{OT}}$ is fixed and the same
for all simulations. The second set of simulations, which generated
\emph{reverse} trajectories, started at time $t=0$ with the optical
trap minimum at $x_{\text{OT}} = x^{\text{final}}_{\text{OT}}$, followed by
the trap minimum being translated with the same set of velocities (but
with opposite sign), until the minimum was located at $x_{\text{OT}} =
0$ at time $t=t_{\text{D}}$. Each simulation in the forward and reverse
direction consisted of $10^5$ trajectories. Ten such simulations were
carried out in each case. The work values obtained after each
trajectory in both sets of forward and reverse simulations (calculated
using Eqn.~(\ref{eq:work})), were sorted into bins of width equal to
0.01. The distributions of work values obtained in this manner are
plotted in Fig.~\ref{fig:crooks} for the various cases.

Panel A in Fig.~\ref{fig:crooks} plots, for parameter set 1, the
probability of work $W$ being performed in the forward path
($P_\text{F}(W)$) alongside the distribution of work values in the
reverse path ($P_{\text{R}}(W)$) for the various trap velocities
$v_{\text{OT}}$ indicated in the figure legend. While the work is
predominantly positive in the forward trajectories (with a positive
mean value), the work is predominantly negative in the reverse
trajectories (with a negative mean value). The widening of the
distributions with increasing trap velocities is also apparent. As
noted previously, in the limit of a quasistatic process ($v_{\text{OT}}
\to 0$), $P_{\text{F}}(W) \to \delta (W - \Delta F)$, and
$\braket{W}_{\text{F}} = \Delta F$. However, for increasing values of
$v_\text{OT}$, the mean value shifts towards the right with a wider
range of work values, and with $\braket{W}_{\text{F}} \ge \Delta F$.

\begin{table*}[t]
\small
\vskip-10pt
\centering
\caption{Comparison of equilibrium free energies calculated
  with the Crooks fluctuation theorem, the Jarzynski equality, and
  from a sum over the first six terms of the cumulant expansion, with
  exact analytical values, for the various trap velocities. The three
  sets of values for the membrane and optical trap potential
  parameters are given in Table \ref{table:par}.}
\label{table:deltaf}
\vskip5pt
\bgroup
\setlength{\tabcolsep}{0.15em}
{\def\arraystretch{1.3}
\begin{tabular}{|c|c|c|c|c|c|c|}
\hline
\multicolumn{7}{|c|}{{\text{Parameter set 1:   $\Delta F_{\text{anal}}=1.796$}}}         \\ \hline    
\hline                                                                           
\multirow{2}{*}{$v_{\text{OT}}$} & \multicolumn{2}{c|}{\text{Crooks}}
& \multicolumn{2}{c|}{\text{Jarzynski} (forward) } 
& \multicolumn{2}{c|}{\text{Cumulants}} \\
\cline{2-7} 
& {$\Delta F$}  & {\% error}  & {$\Delta F$}   & {\% error}    
& {$\Delta F_6$}          & {\% error}          \\
\hline
\multicolumn{1}{|l|}{0.01}  &  $1.80 \pm 0.02$ & 0.22 &   $1.7955 \pm 0.0002$ & 0.03  
&         1.796           &      0.03              \\
\hline
\multicolumn{1}{|l|}{0.05}  & $1.79 \pm 0.02$ & 0.34  & $1.796  \pm 0.001$    &  0.004          
&      1.799              &   0.14                 \\
\hline
\multicolumn{1}{|l|}{0.1}  &   $1.80 \pm 0.03 $ & 0.22 &   $1.796  \pm 0.002$  & 0.03        
&     1.797               &      0.06             \\
\hline
\multicolumn{1}{|l|}{0.5}   &    $1.76 \pm 0.05$ & 2.01  & $1.808  \pm 0.016$ & 0.65              
&        1.823            &      1.48              \\
\hline
\multicolumn{1}{|l|}{1}                 &   $1.81 \pm 0.05$ & 0.78  &
 $1.834       \pm 0.044$           &        2.12            
&      1.746              &     2.81               \\
\hline
\hline
\multicolumn{7}{|c|}{{\text{Parameter set 2: $\Delta F_{\text{anal}}=7.960$}}}                                   \\ 
\hline
\hline
\multirow{2}{*}{$v_{\text{OT}}$} & \multicolumn{2}{c|}{\text{Crooks}} 
& \multicolumn{2}{c|}{\text{Jarzynski} (forward) } 
& \multicolumn{2}{c|}{\text{Cumulants}} \\ 
\cline{2-7} 
& {$\Delta F$}  & {\% error}  & {$\Delta F$}          & {\% error}          
& {$\Delta F_6$}          & {\% error}          \\ 
\hline
\multicolumn{1}{|l|}{0.01}                  &    $7.96 \pm 0.04$ & 0.004  &
  $7.9600 \pm 0.0005$             &        0.003            
&       7.961             &     0.01               \\ 
\hline
\multicolumn{1}{|l|}{0.05}                  &    $7.96 \pm 0.05$ & 0.004  &
    $7.960\pm 0.001 $        &        0.01            &        7.964            
&      0.05              \\
\hline
\multicolumn{1}{|l|}{0.1}                  &     $7.96 \pm 0.04$ & 0.004 &
  $7.962 \pm 0.002$                &        0.02            
&       7.951             &   0.12                 \\ 
\hline
\multicolumn{1}{|l|}{0.5}                  &      $7.88 \pm 0.09$ & 1.01  &
     $7.974 \pm 0.023$              &       0.17             
&           8.019         &          0.74          \\ 
\hline
\multicolumn{1}{|l|}{1}                  &     $7.96 \pm 0.09$ & 0.004 &
  $8.165   \pm 0.035$             &       2.57             
&       8.213            &    3.18                \\ 
\hline
\hline
\multicolumn{7}{|c|}{{\text{Parameter set 3:  $\Delta F_{\text{anal}}=0.934$}}}                                  \\ 
\hline
\hline
\multirow{2}{*}{$v_{\text{OT}}$} & \multicolumn{2}{c|}{\text{Crooks}} 
& \multicolumn{2}{c|}{\text{Jarzynski} (forward) } 
& \multicolumn{2}{c|}{\text{Cumulants}} 
\\ 
\cline{2-7} 
                  & {$\Delta F$}  & {\% error}  & {$\Delta F$}          
& {\% error}          & {$\Delta F_6$}          & {\% error}          
\\ 
\hline
\multicolumn{1}{|l|}{0.01}                  &   $0.94 \pm 0.02$ & 0.69  &
 $0.9333 \pm 0.0002$               &        0.02            
&      0.933              &     0.05            
\\ 
\hline
\multicolumn{1}{|l|}{0.05}                  
&       $0.94 \pm 0.03$ & 0.69  &
    $0.934  \pm 0.001 $             &          0.03         
&        0.937            &      0.34              
\\
\hline
\multicolumn{1}{|l|}{0.1}                  
&      $0.93 \pm 0.04$ & 0.38  &
     $0.933  \pm 0.001$             &        0.02           
&       0.933             &         0.08          
\\ 
\hline
\multicolumn{1}{|l|}{0.5}                  
&       $0.93 \pm 0.04$ & 0.38   &
     $ 0.936  \pm 0.012 $           &          0.21          
&          0.955          &        2.25            
\\ 
\hline
\multicolumn{1}{|l|}{1}                  
&       $0.94 \pm 0.06$ & 0.69   &
   $0.933     \pm 0.020 $           &      0.03             
&       1.063             &     13.91               \\ \hline
\end{tabular}
}
\egroup
\vspace{-10pt}
\end{table*}

The usefulness of Crooks fluctuation theorem is best appreciated when
$P_{\text{F}}(W)$ is plotted alongside $P_{\text{R}}(- W)$ as shown in
panels B, C and D of Fig.~\ref{fig:crooks}. These three figure panels
correspond to the three potential parameter sets listed in Table
\ref{table:par}, respectively. As noted before, according to
Eqn.~(\ref{eq:FTcft}), the value of work $W^*$ at which
$P_{\text{F}}(W^*) = P_{\text{R}}(- W^*)$ is nothing but the equilibrium
free energy difference. Consequently, $\Delta F$ is estimated from
Fig.~\ref{fig:crooks} by finding the point of intersection of the
forward and reverse probability curves for each of the trap
velocities, for the three sets of parameter values. The values of
$\Delta F$ obtained in this way are listed in
Table~\ref{table:deltaf}, along with an estimate of the error in 
finding the point of intersection due to the relatively coarse 
interval used for binning the work values. The percentage 
relative error in the free energy predicted by the Crooks 
fluctuation theorem, defined by the expression
\begin{equation}
  \text{Error} = \left\vert \frac{\Delta F -
    \Delta F_{\text{anal}}}{\Delta F_{\text{anal}}}
    \right\vert \times 100
\label{eq:err}
\end{equation}
is also listed in Table~\ref{table:deltaf}. It is worth noting that the error in finding the point of intersection consistently increases with the trap velocities, but is roughly the same order of magnitude in all cases. On the other hand, the percentage relative error varies without a set pattern for the different values of $v_{\text{OT}}$, depending on how close the predicted value is to the analytical value. Remarkably, for each parameter set, the intersection of the forward and reverse probability curves occurs at nearly identical values, with the error in the
estimated free energy being at most $2\%$ even for large trap
velocities.

The increase in error with increasing trap velocity can be understood
by considering panel B in Fig.~\ref{fig:crooks}. As the velocity
increases, it causes the mean value of work to shift away from the
free energy value, with a simultaneous increase in the standard
derivation of the distribution. As a result, the crossover occurs at
the tails of the distributions, where errors are high and therefore
require much larger populations to ensure adequate
statistics. Figure~\ref{fig:crooks} indicates that the velocities at
which this could become an issue is sensitive to the choice of
potential parameters. Parameter set~1 (panel B), where the optical
trap strength was double that of the membrane, and the barrier height
for detachment was much lower than that of re-attachment (see
Fig.~\ref{fig:par}a), seems to have the most movement of the mean away
from the exact free energy value. On the other hand parameter set 3
(panel D), where barrier heights are of ${\mathcal O} (k_{\text{B}}T)$
(see Fig.~\ref{fig:par}c), seems to be the least affected by increased
velocity.

\subsection{Jarzynski equality}

The form of the Jarzynski equality given by
Eqn.~(\ref{eq:FTje}) corresponds to switching the system from an
initial equilibrium state with $\lambda = \lambda_0$ to a final state
with $\lambda = \lambda_{\text{f}}$. When the system is switched
from an initial equilibrium state with $\lambda = \lambda_{\text{f}}$ to
a final state with $\lambda = \lambda_0$, the Jarzynski equality takes the form~\cite{Hummer:2001ec},
 \begin{equation}
 \braket{e^{- W}}_\text{R} = e^{ \Delta F}
 \label{eq:FTje2}
 \end{equation}
where the subscript `R' on the ensemble average on the left hand side
indicates an average over reverse trajectories, and the change in free
energy is still defined by $\Delta F = F_{\lambda_\text{f}} -
F_{\lambda_0}$.

\begin{figure*}[!ht]
\centering
\includegraphics[width=18cm]{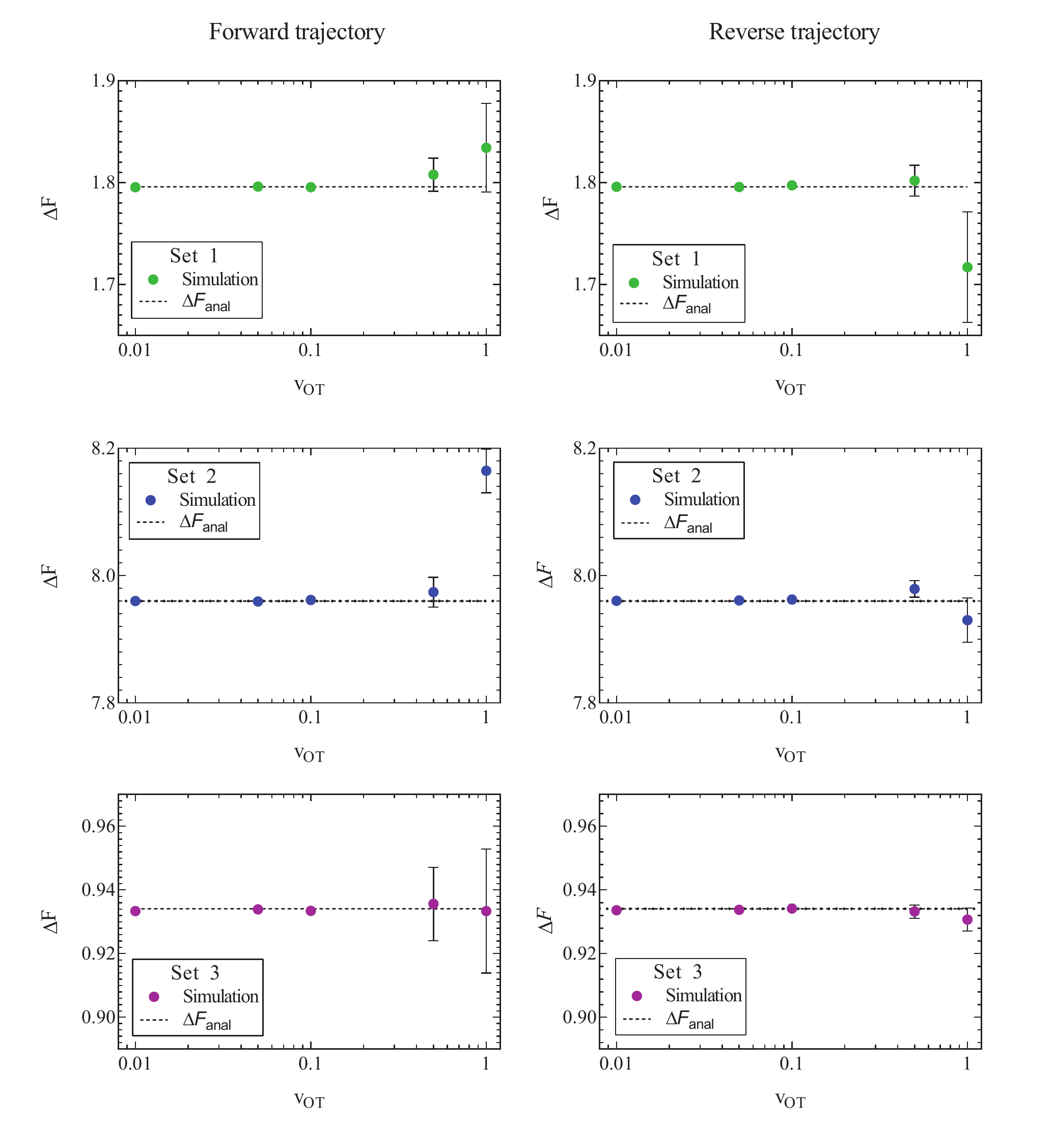}
\caption{Free energy values estimated using Jarzynski's equality as a
  function of trap velocity $v_{\text{OT}}$. Symbols are results of
  simulations, while the dashed lines indicate the exact analytical
  value of the free energy, for parameter sets 1 (row 1), 2 (row 2)
  and 3 (row 3). Results for the forward trajectories are displayed in
  column one, whilst reverse trajectories are displayed in column
  two. Error bars indicate the standard error in the estimated mean
  free energy values obtained from ten repeated simulations.}
\label{fig:je2}
\end{figure*}

The sets of forward and reverse simulations carried out to demonstrate
the Crooks fluctuation theorem can also be used to examine the
usefulness of the Jarzynski equality. The ensemble averages on the
left hand sides of Eqns.~(\ref{eq:FTje}) and~(\ref{eq:FTje2}) were
calculated using the values of work accumulated at the end of each of
the $10^5$ trajectories corresponding to a particular simulation. The
sets of forward and reverse simulations were repeated ten times each,
so that we obtain ten estimates for the equilibrium free energy in
each case, and the errors can be estimated. The mean of these 10
values, and the standard error in these mean values are displayed in
Fig.~\ref{fig:je2} for all the cases considered here. Parameter sets
1, 2, and 3 are shown in rows 1, 2, and 3 respectively, with the left
hand column showing results for the forward trajectories whilst the
right hand column shows results for reverse trajectories. The mean
value of $\Delta F$ and the standard error in the mean are also
compared with exact analytical values in Table~\ref{table:deltaf} for
simulations carried out in the forward direction. Note that the
percentage relative error reported in the Table is calculated using
Eqn.~({\ref{eq:err}) with the mean value of $\Delta F$.
 
A feature of all approaches for determining free energy differences
using ensemble averages, of which the Jarzynski equality is no
exception, is their limitation due to sample size. As argued by
Jarzynski~\cite{Jarzynski:1997fu}, for systems where the spread in the
distributions $P_\text{F}(W)$ and $P_\text{R}(W)$ is large, the
function $\exp(-W)$ varies significantly over many standard deviations
about the mean value of work. As a result, the numerically determined
average $\braket{\exp(-W)}$ can be dominated by work values that are
by their very nature statistically rare. Therefore an unreasonable
number of measurements of the work would be required to get an
accurate result. This results in a practical restriction on the rates
at which the system can be switched between $\lambda_0$ and
$\lambda_\text{f}$. As can be seen from Fig.~\ref{fig:je2} and
Table~\ref{table:deltaf}, the accuracy in the estimation of the free
energy decreases with the trap velocity in all cases.
\begin{table}[t]
\centering
\caption{Accuracy of the Gaussian approximation at various trap
  velocities in the forward and reverse paths, for the membrane and
  optical trap potential parameters corresponding to Set 1 in
  Table~\ref{table:par} }
\label{table:gauss}
\bgroup
\setlength{\tabcolsep}{0.3em}
{\def\arraystretch{1.3}
\begin{tabular}[t]{| c | c | c | c | c | c | c |}
\hline
\multicolumn{6}{|c|}{{\text{Forward trajectories}}}  
\\ 
\hline
\hline 
\multicolumn{1}{|l|}{$v_{\text{OT}}$} & $\Delta F_{\text{F}}$ 
&    $\braket{W}_{\text{F}}$ & $\sigma^2_{\text{F}}$ &  $\braket{W_d}_{\text{F}}$  
&    $E_{\text{F}}$  
\\ 
\hline                                                                            
\multicolumn{1}{|l|}{0.01} &  \multirow{5}{*}{1.796} &    1.860 &   0.128 
& 0.064  &   0.000                     
\\ 
\cline{1-1} \cline{3-6} 
\multicolumn{1}{|l|}{0.05} &    &    2.116 &   0.632 & 0.320   &   0.003      
\\ 
\cline{1-1} \cline{3-6}
\multicolumn{1}{|l|}{0.1} &    &    2.428 &   1.262 & 0.632   &   0.001      
\\ \cline{1-1} \cline{3-6}
\multicolumn{1}{|l|}{0.5}      &            &        4.842         &      5.922 
& 3.046         &        0.085     
\\ 
\cline{1-1} \cline{3-6}
\multicolumn{1}{|l|}{1}   &                 &               7.535            
&      10.604  & 5.739      &      0.437 
\\ \hline \hline
\multicolumn{6}{|c|}{{\text{Reverse trajectories}}}         
\\ \hline   
\hline
\multicolumn{1}{|l|}{$v_{\text{OT}}$} & $\Delta F_{{\text{R}}}$ 
&    $\braket{W}_{\text{R}}$ & $\sigma^2_{\text{R}}$ &  $\braket{W_d}_{\text{R}}$  
&    $E_{\text{R}}$   
\\ 
\hline                                                                            
\multicolumn{1}{|l|}{0.01} &  \multirow{5}{*}{-1.796} &    -1.732 
&   0.128 & 0.064  &        0.000
\\ 
\cline{1-1}  \cline{3-6}
\multicolumn{1}{|l|}{0.05} &    &    -1.481 &   0.633 & 0.315   
&   0.001                     
\\
\cline{1-1} \cline{3-6}
\multicolumn{1}{|l|}{0.1} &    &    -1.159 &   1.281 
& 0.637   &   0.003                     
\\ 
\cline{1-1} \cline{3-6}
\multicolumn{1}{|l|}{0.5}      &            &      1.320          
&      6.337 &  3.116  & 0.052
\\ 
\cline{1-1} \cline{3-6}
\multicolumn{1}{|l|}{1}   &               &        4.258 
& 12.552 & 6.054   & 0.222
\\ 
\hline 
\end{tabular}
}
\egroup
\end{table}

A comparison of the relative errors in the free energies predicted by
the Crooks fluctuation theorem and the Jarzynski equality (in the case
of forward trajectories) in Table~\ref{table:deltaf} shows that they
are roughly similar in magnitude for the various cases. As noted
earlier, there is a reduction in accuracy with increasing trap
velocity, which appears to be magnified when either one or both the
potential well depths are high compared to $k_{\text{B}}T$, which is
the case for parameter sets 1 and 2 (displayed in
Fig.~\ref{fig:par}). The dependence of the error on well depth is
studied shortly below.

For slow rates of switching between $\lambda_0$ and
$\lambda_{\text{f}}$, the distributions $P_{\text{F}}(W)$ and
$P_{\text{R}}(W)$ are expected to be approximately
Gaussian~\cite{Hummer:2001ec}. In this case, retaining only the first
two terms in the cumulant expansion for $\braket{\exp(-W)}$ (which is
discussed in greater detail in the section below), one can
write~\cite{Hummer:2001ec},
\begin{align}
\Delta F_{\text{F}} & = F_{\lambda_\text{f}} - F_{\lambda_0} 
\approx \braket{W}_{\text{F}} - \frac{\sigma^2_{\text{F}}}{2}   \nonumber \\
\Delta F_{\text{R}} & = F_{\lambda_0} - F_{\lambda_\text{f}} \approx 
\braket{W}_{\text{R}} - \frac{\sigma^2_{\text{R}}}{2}   \nonumber
\end{align}
where $\sigma^2_{\text{F}}$ and $\sigma^2_{\text{R}}$ are the variances of
the work distributions $P_{\text{F}}(W)$ and $P_{\text{R}}(W)$,
respectively. Defining the mean dissipated work $\braket{W_d}$ as the
difference between the mean actual work of the process and the
reversible work (which is equal to the equilibrium free energy), we
can estimate the departure from the Gaussian approximation by
evaluating the error estimates $E_{\text{F}}$ and $E_{\text{R}}$ defined
by,
\begin{align}
  E_{\text{F}} & =  \left[ \braket{W}_{\text{F}} - \frac{\sigma^2_{\text{F}}}{2}
    \right] - \Delta F_{\text{F} }
= \braket{W_d}_{\text{F}} - \frac{\sigma^2_{\text{F}}}{2} 
\\
E_{\text{R}} & =  \left[ \braket{W}_{\text{R}} -
  \frac{\sigma^2_{\text{R}}}{2} \right] 
- \Delta F_{\text{R}}
= \braket{W_d}_{\text{R}} - \frac{\sigma^2_{\text{R}}}{2} 
\end{align}

The values of mean actual work, variances, mean dissipated work and
error estimates, for membrane and optical trap potential parameters
corresponding to Set 1, are displayed in Table~\ref{table:gauss} for
both the forward and reverse paths. Clearly, the Gaussian
approximation leads to an error of less than $9\%$ up to trap
velocities $v_{\text{OT}} = 0.5$. Interestingly, the variances of
$P_{\text{F}}(W)$ and $P_{\text{R}}(W)$ and the mean dissipated work in
the forward and reverse paths are roughly equal in magnitude for
identical velocities in the forward and reverse paths.

For distributions that are not Gaussian, the exponential average in
Jarzynski's equality can be expanded in terms of
cumulants~\cite{Hummer:2001ec}, and the convergence of $\Delta F$ can
be studied as a function of the various potential parameters, as
discussed in the section below. It is worth noting that it is also
possible to obtain estimates for the free energy that are accurate to
a higher order in the cumulant expansion than the Gaussian
approximation by suitably combining the mean work and variance in the
forward and reverse paths~\cite{Hummer:2001ec}.

\subsection{Cumulant expansion for the free energy of binding}
\label{cumul}

The average of the exponential of work on the left hand sides of
Eqns.~(\ref{eq:FTje}) and~(\ref{eq:FTje2}) in Jarzynski's equality can
be expanded in terms of cumulants\cite{Hummer:2001ec}. In the case of
forward paths, this leads to the following expression for the free
energy change:
\begin{equation}
\Delta F = \lim_{k \to \infty}  \Delta F_k ,
\label{feexp}
\end{equation}
where
\begin{equation}
\Delta F_k = \sum_{n=1}^{k} (-1)^{n+1} \, \frac{C_n}{n!} .
\label{cumexp}
\end{equation}
Here, the cumulants $C_n$ are defined by the expressions
\begin{align}
\label{mom2cu}
C_1&=\braket{W}_{\text{F}} \nonumber \\
C_2 &=\mu_2 = \sigma_{\text{F}}^2 \nonumber \\
C_3&=\mu_3 \nonumber\\
C_4&=\mu_4 - 3 \, \mu_2^2 \nonumber \\
C_5&=\mu_5 - 10  \,\mu_2  \,\mu_3 \nonumber\\
C_6&=\mu_6 - 15  \,\mu_2 \,\mu_4 - 10  \,\mu_3^2 + 30  \,\mu_2^3 \nonumber\\[-5pt]
\vdots \nonumber\\[-5pt]
C_n & = \mu_n - \sum_{j=1}^{n-2} \binom{n-1}{j}  \,\mu_j  \,C_{n-j} \, ;
\quad n \ge 2
\end{align}
with $\mu_n$ being the central moments of $P_{\text{F}}(W)$,
\begin{equation}
  \mu_n = \left\langle \, \left[ W - \braket{W}_{\text{F}} \right]^n \,
  \right\rangle_{\text{F}} .
\label{mom}
\end{equation}
The recursive relationship between the cumulants and central moments
in Eqn.~(\ref{mom2cu}) has been given by Smith~\cite{Smith:1995hr}. In
the case of reverse paths, the cumulant expansion on the right hand
side of Eqn.~(\ref{cumexp}) leads to the free energy change $ - \Delta
F = F_{\lambda_0} - F_{\lambda_\text{f}}$, with $\mu_n$ in the
expressions for $C_n$ being the central moments of $P_\text{R}(W)$.

\begin{figure}[!t]
\centering
\includegraphics[width=9 cm]{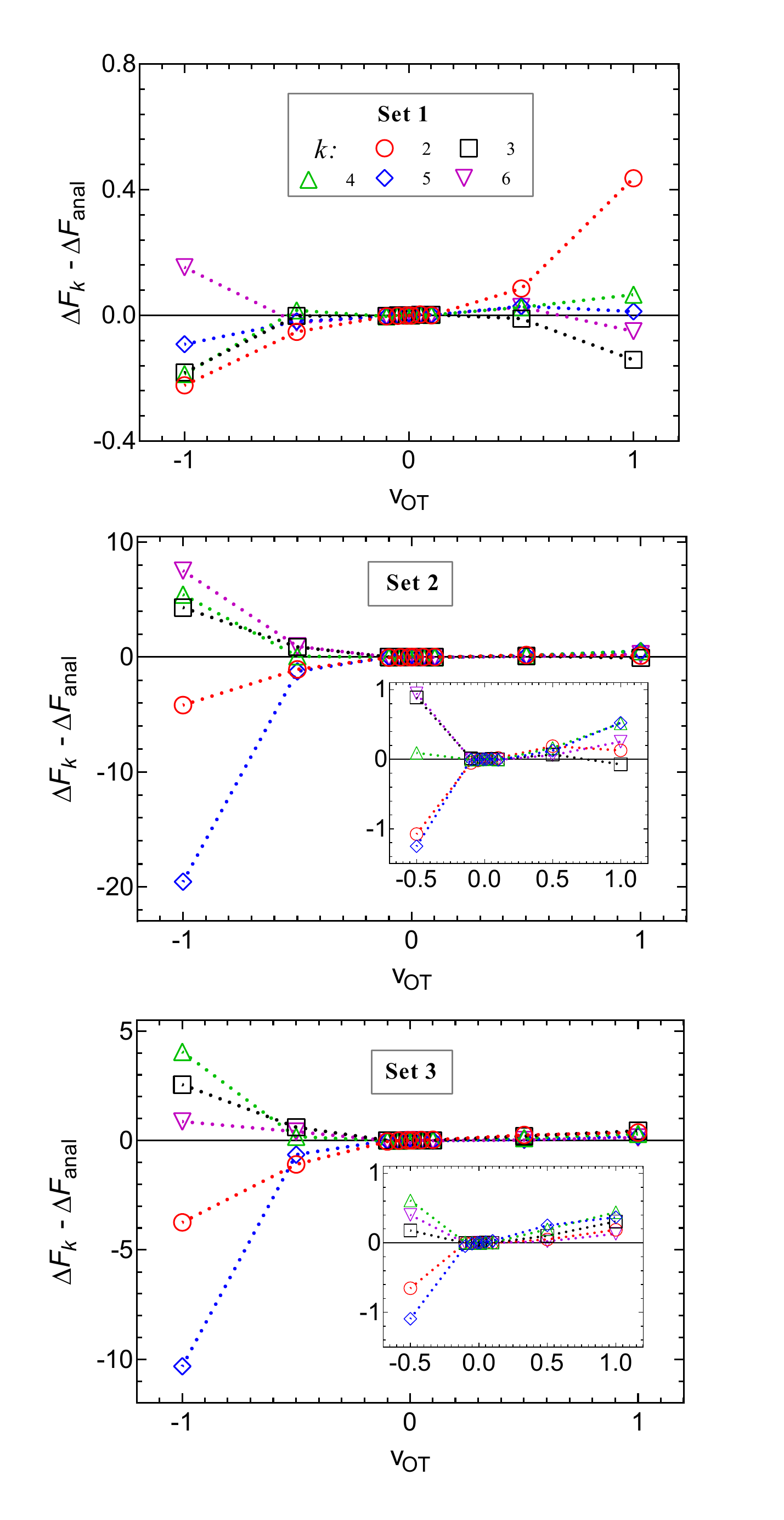}
\caption{Deviation of the approximate estimate of the free 
energy change $\Delta F_k$, obtained from a cumulant expansion, 
from the analytical free energy $\Delta F_{\text{anal}}$, at  various 
values of trap velocity $v_\text{OT}$, for different numbers of terms $k$ in the
  expansion. Symbols are results of simulations for
  parameter sets 1 (row 1), 2 (row 2) and 3 (row 3). 
  Lines are drawn to guide the eye. Data for the
  forward trajectories and reverse trajectories are displayed together 
  by representing the velocities in the latter case with negative values. 
  The insets for Sets 2 and 3 make it easier to identify the values of $\Delta F_k- \Delta F_{\text{anal}}$ for all values of $v_\text{OT} \neq -1$. }
\label{fig:cum1}
\end{figure}

An analysis of the simulation results for the forward and reverse
paths in terms of the cumulant expansion is displayed in
Fig.~\ref{fig:cum1}, where the difference between the values of $\Delta F_k$ (which
represent the approximate estimate of the free energy change given by
$k$ terms of the cumulant expansion) and the analytical value $\Delta
F_{\text{anal}}$, is plotted against the trap velocities $v_{\text{OT}}$ (for  
values of $k$ in the range $ 2 \le k  \le 6$). Additionally, the particular values obtained for
$\Delta F_6$ in the case of forward trajectories, and the relative
error compared to the exact values are listed in
Table~\ref{table:deltaf}. As expected, at low trap velocities where
the system approaches a quasistatic process, the work distribution
approaches a Gaussian, and quite accurate results are obtained with
two cumulants. However as the trap velocity increases, higher cumulant
numbers are required until, for $v_{\text{OT}} = 1$, even at cumulant
numbers of 6 the system has still not converged. 

An alternative representation of the cumulant expansion
 data is given in Fig.~\ref{fig:cum3}, where 
$\Delta F_k- \Delta F_{\text{anal}}$ is plotted as a function of $k$ ($ 2 \le k  \le 6$), at the 
lowest and highest  trap velocities ($v_{\text{OT}} = 0.01$ and $v_{\text{OT}} = 1.0$), for parameter values corresponding to set 3. Since the cumulant expansion is an approximation for the left hand sides of Eqns.~(\ref{eq:FTje}) and~(\ref{eq:FTje2}), we expect that the free energy difference $\Delta F_k$ should converge to the free energy difference predicted by Jarzynski's equality $\Delta F_\text{Jarzynski}$, for sufficiently large values of $k$. This can be seen to be clearly the case for $v_{\text{OT}} = 0.01$, for both the forward and reverse trajectories, from the top row in Fig.~\ref{fig:cum3}, where the solid line corresponds to the difference $\Delta F_\text{Jarzynski}-  \Delta F_{\text{anal}}$. The scale of the $y$-axis in both the subfigures in the bottom row of Fig.~\ref{fig:cum3} (corresponding to $v_{\text{OT}} = 1.0$) makes it difficult to distinguish $\Delta F_\text{Jarzynski}-  \Delta F_{\text{anal}}$ from 0. While the values of $\Delta F_k- \Delta F_{\text{anal}}$ appear to be getting smaller with increasing $k$, there are still large changes in $\Delta F_k$ with increasing $k$, and convergence has not occurred by $k=6$, as was observed previously at this value of trap velocity in Fig.~\ref{fig:cum1}. 

The cumulant expansion can also be used to examine the influence of
well depth. In order to do so, simulations in the forward direction
were carried out for $10^6$ trajectories with time step $\Delta t =
10^{-4}$, for trap velocities $v_{\text{OT}} = \{0.01, 0.05, 0.1, 0.5,
1\}$. In all cases, the final location of the trap potential minimum
was $x_{\text{OT}}^{\text{final}} = 6$. The membrane potential depth
was held fixed at $\epsilon_{\text{M}} = 4$, whilst a parameter sweep
from 1 to 8 was carried out for the optical trap potential depth,
$\epsilon_{\text{OT}}$. The trap strengths $k_{\text{M}}$ and
$k_{\text{OT}}$ for both the membrane and the optical trap potentials
were held constant at a value of two. Results of the cumulant analysis
are plotted in Fig.~\ref{fig:cum2} for the difference $\Delta F_k- \Delta F_{\text{anal}}$, as a function of trap velocity, at the various values of $k$,
with each subfigure representing a different value of
$\epsilon_{\text{OT}}$. Since the exact analytical value $\Delta
F_{\text{anal}}$ is different for each value of trap well depth, 
the values are given in the caption to Fig.~\ref{fig:cum2}.
\begin{figure*}[t]
\centering
\includegraphics[width=18 cm]{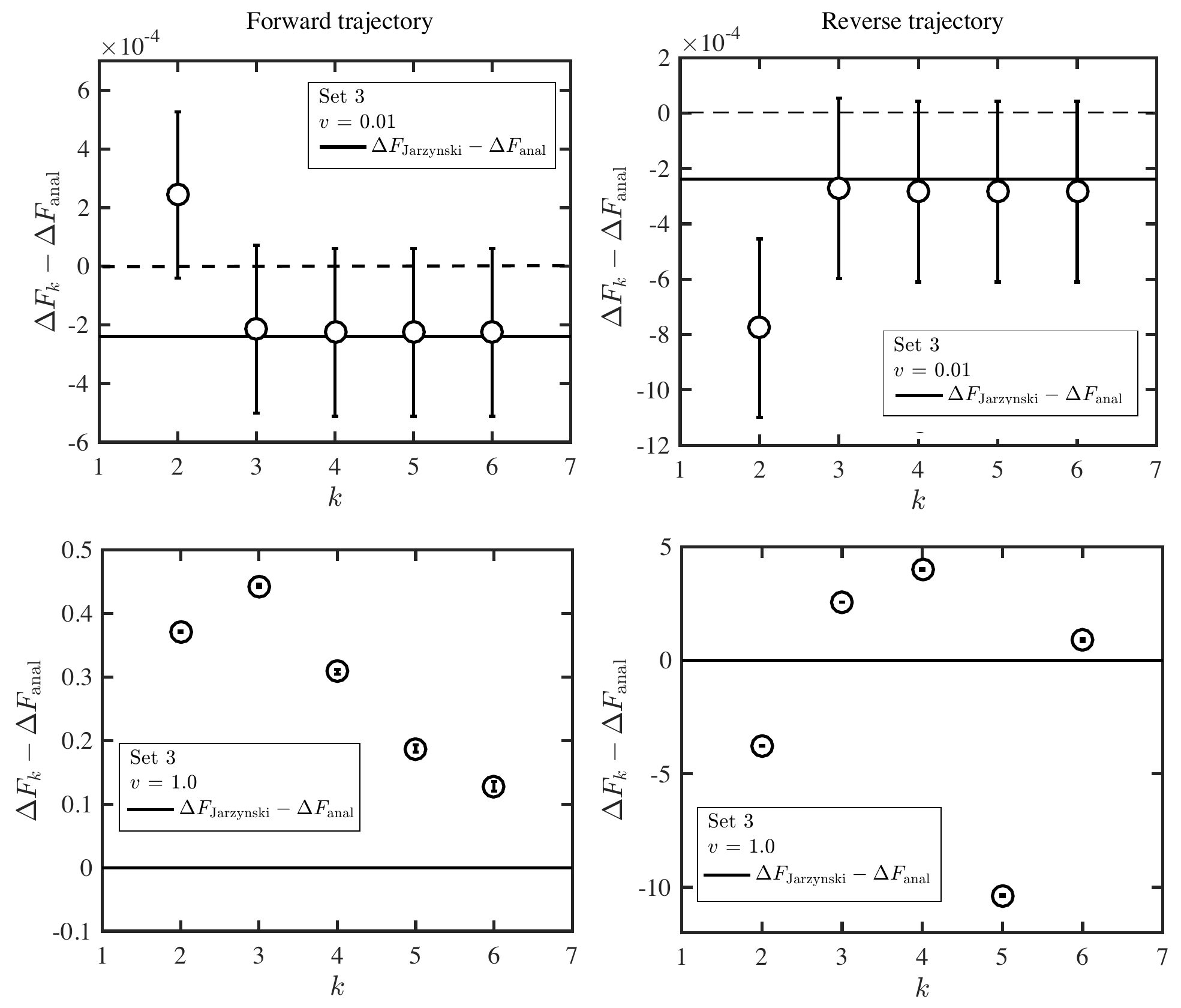}
\caption{Deviation of the approximate estimate of the free 
energy change $\Delta F_k$, obtained from a cumulant expansion, 
from the analytical free energy $\Delta F_{\text{anal}}$, as a function of the  
numbers of terms $k$ in the expansion, at two values of 
the trap velocity $v_\text{OT}$, for parameter values corresponding to set 3. 
The full lines indicate the difference $\Delta F_\text{Jarzynski}-  \Delta F_{\text{anal}}$. 
Results for the forward trajectories are displayed in column one, whilst reverse
  trajectories are displayed in column two. }
\label{fig:cum3}
\end{figure*}
The cumulant analysis suggests that convergence occurs quickly at the
low velocities and becomes poorer and poorer at higher velocities. It is also evident that
increasing optical trap well depth significantly increases the error
in the estimate of the free energy for a given value of the number of
terms $k$ in the cumulant expansion (note the different scales of the $y$-axes in the different subfigures of Fig.~\ref{fig:cum2}).

\subsection{Probabilities of attachment and detachment via umbrella sampling}
 
An important quantity that is frequently the focus of experiments on
cell adhesion is the probability of adhesion. Measurements of the
adhesion probability are often used to determine the kinetics of the
adhesion process through the calculation of on and off-rates of
binding etc. The experiments, which typically monitor whether a
binding event occurs or not when ligand and receptor bearing surfaces
are brought into contact, are by their very nature carried out at
finite rates. As a result, a true measure of the equilibrium
probability of binding is difficult to obtain. In this context, the
method of non-equilibrium umbrella
sampling~\cite{Crooks:2000ke,Williams:2011ei,williams2010Non,gao2012non}
provides a means of determining the equilibrium binding probability
from non-equilibrium measurements. Here, we demonstrate how
non-equilibrium umbrella sampling can be used to find, at the end of
the unbinding experiment, the probability of either the bead being
attached to the cell, or being detached from it and held in the
optical trap.
\begin{figure*}[!ht]
\begin{center}
\begin{tabular}{cc}
\includegraphics[width=16.5cm]{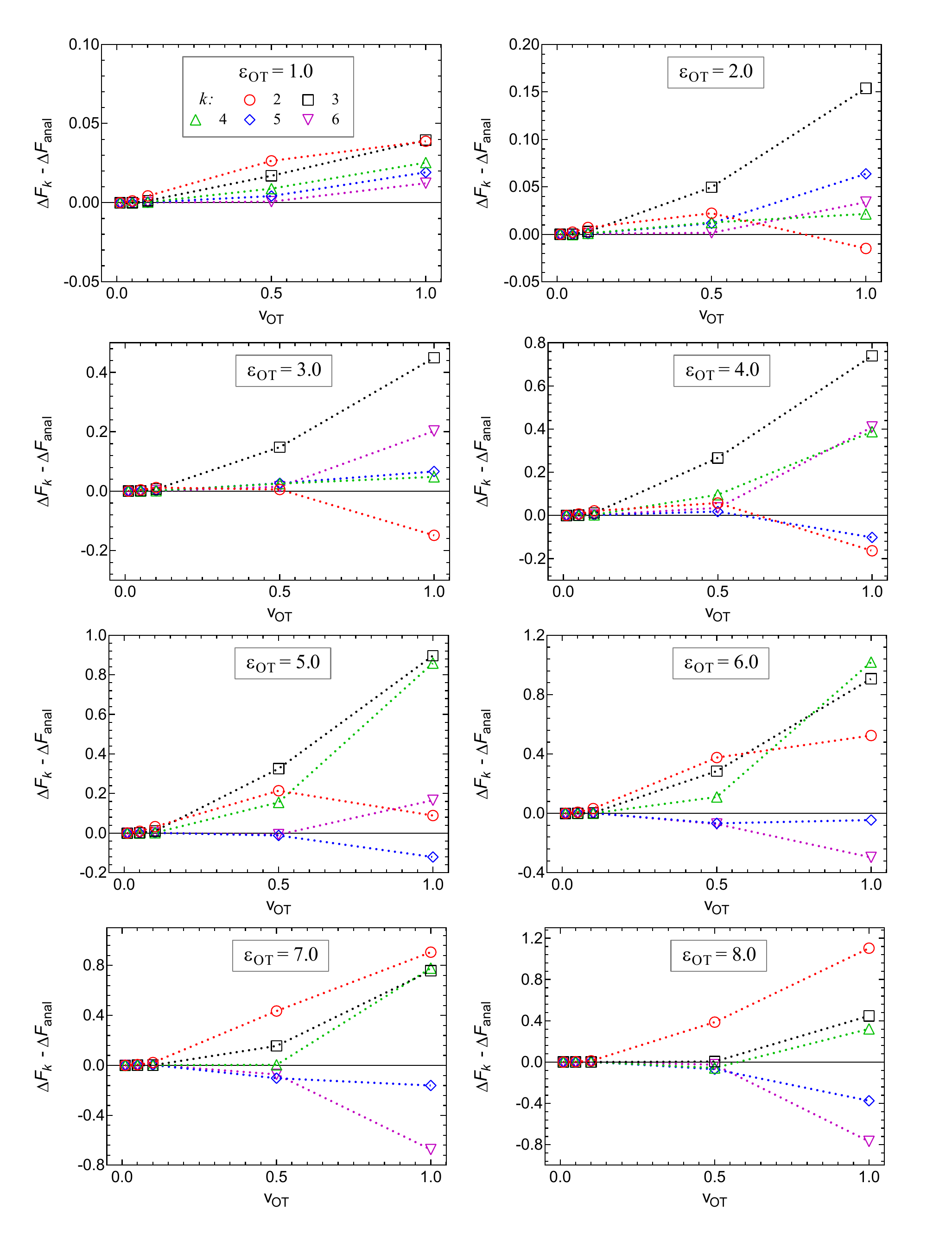}
\end{tabular}
\end{center}
\vskip-15pt
\caption{Influence of the optical trap well depth on 
$\Delta F_k-\Delta F_{\text{anal}}$, for  $ 2 \le k  \le 6$, calculated at  
various values of trap velocity $v_\text{OT}$.  
A parametric sweep was carried out from $\epsilon_{\text{OT}} =1$ (top
  left) to 8 (bottom right), whilst keeping all other potential
  parameters constant ($\epsilon_{\text{M}} = 4$, $k_{\text{M}} =2$ and
  $k_{\text{OT}} = 2$). The exact analytical values of the free energy
  for each of the optical trap depths were, $(\epsilon_{\text{OT}},
  \Delta F_{\text{anal}})$: $(1.0, 0.599574)$, $(2.0, 1.509950)$, $(3.0,
  2.327020)$, $(4.0, 2.952370)$, $(5.0, 3.336500)$, $(6.0, 3.525130)$,
  $(7.0, 3.604400)$, $(8.0, 3.635160)$. }
\label{fig:cum2}
\end{figure*}

At the end of the computer experiment, when $t = t_{\text{D}}$ and the
optical trap minimum is located at $x_{\text{OT}}^{\text{final}}$, it
makes sense to sub--divide the $x$ axis into \emph{three} intervals
(cf. Fig.~\ref{fig:timepot} (c)): First, there is the interval $-
\infty < x(t_{\text{D}}) \le x_{\text{M}}^{\text{ub}}$, which we
define as the set of states that correspond to the bead still being
\emph{attached} to the cell (membrane). The second interval is
$x_{\text{M}}^{\text{ub}} < x(t_{\text{D}}) <
x_{\text{OT}}^{\text{lb}}$, where the potential is flat, and which we
define as corresponding to an \emph{intermediate} state of the
bead. Finally, the interval $x_{\text{OT}}^{\text{lb}} \le
x(t_{\text{D}}) < \infty$ corresponds, according to our definition, to
the \emph{detached} (or optically trapped) state of the bead. In what
follows, we will focus on the equilibrium probabilities for the
attached state and the detached state; the probability for the
intermediate state then follows trivially by subtracting the sum of
these values from one.

\begin{figure*}[!ht]
\begin{center}
\begin{tabular}{cc}
\resizebox{8.25cm}{!} {\includegraphics*[width=12cm]{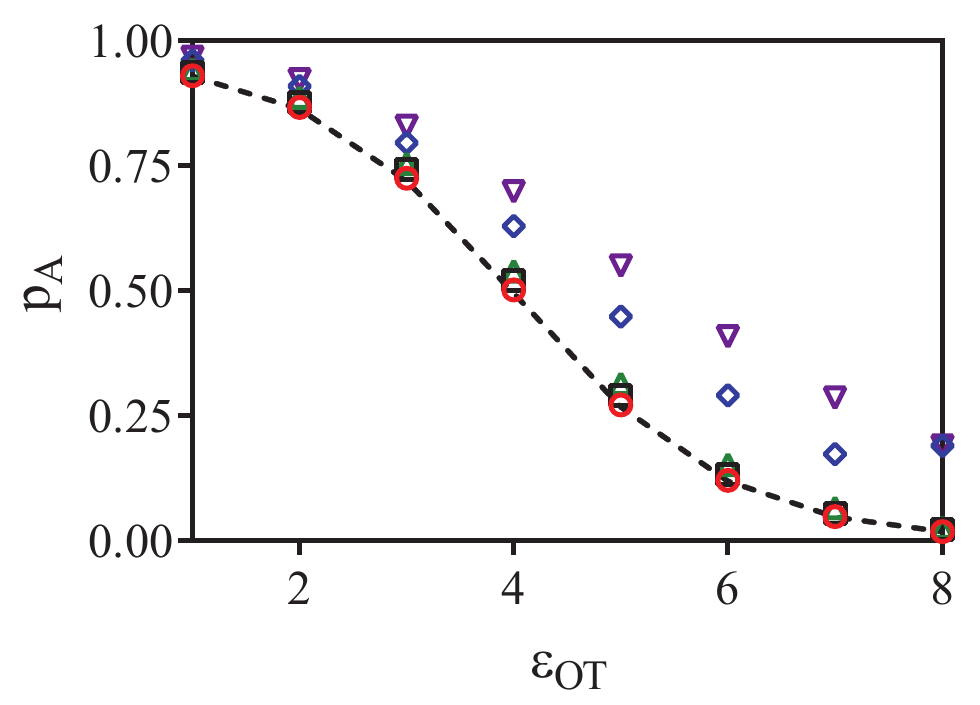}} & \hskip-5pt
\resizebox{8.35cm}{!} {\includegraphics*[width=12cm]{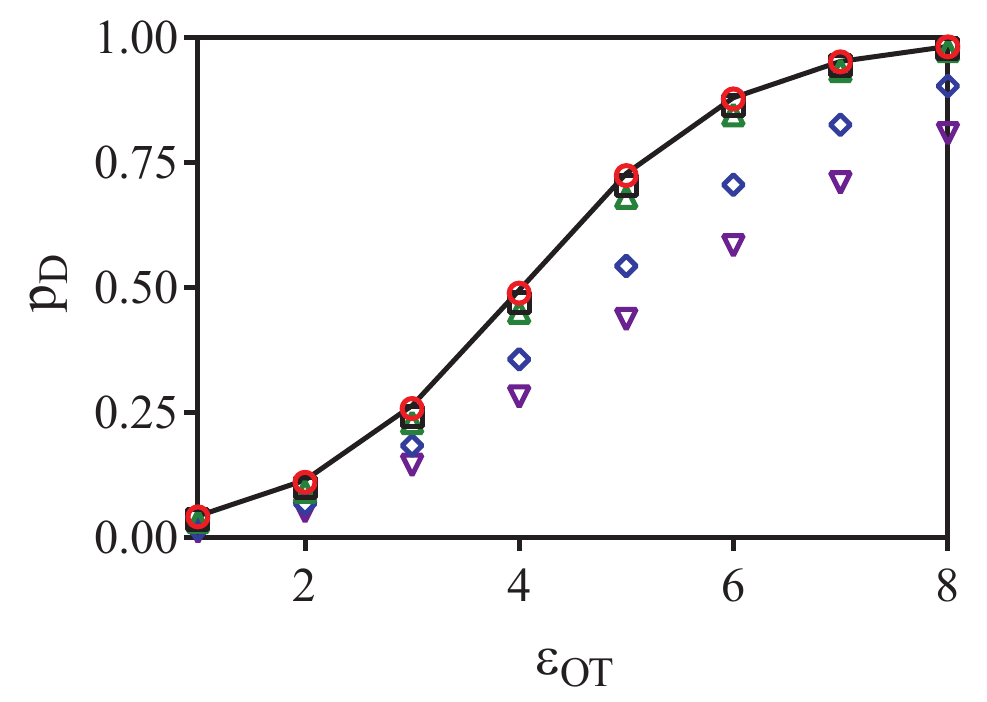}}\\
(a) & (b) \\
\resizebox{8.15cm}{!} {\includegraphics*[width=12cm]{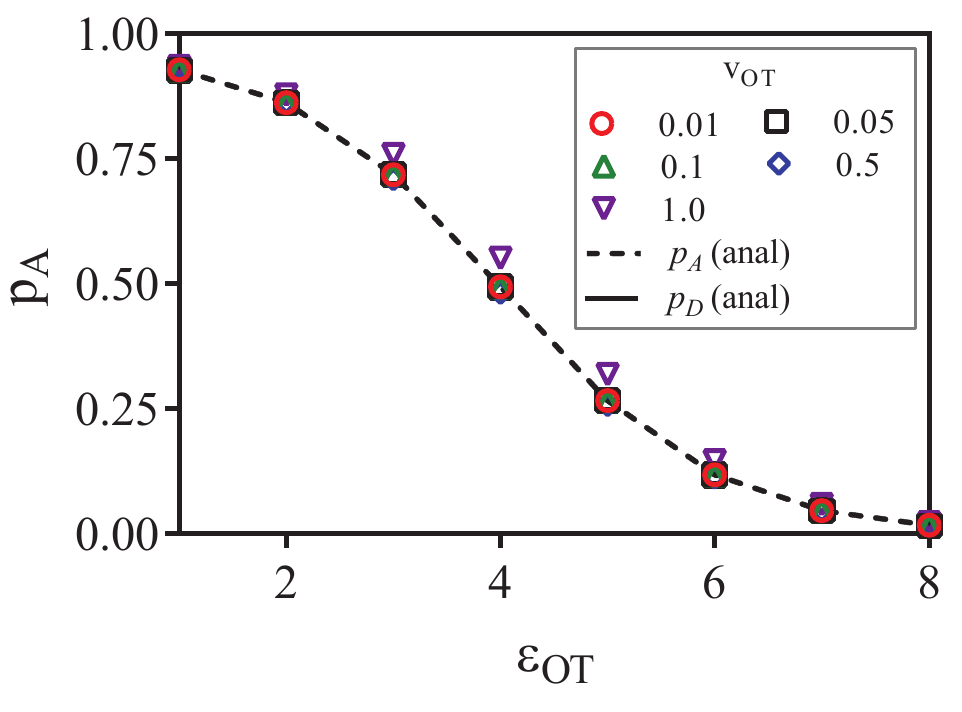}} & \hskip-5pt
\resizebox{8.45cm}{!} {\includegraphics*[width=12cm]{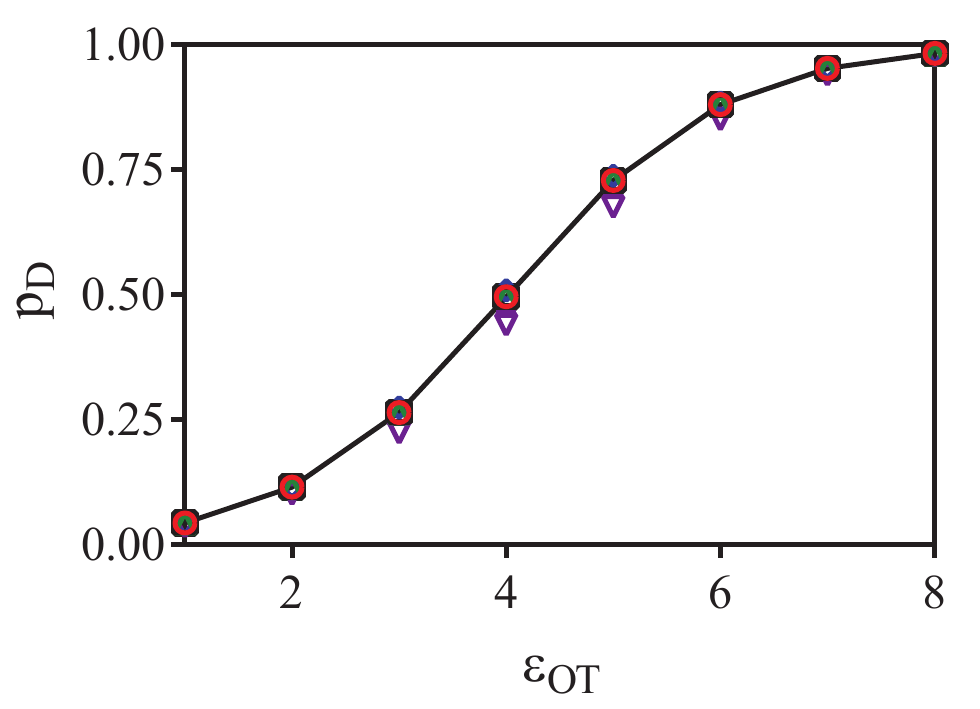}}\\
(c) & (d) \\
\end{tabular}
\end{center}
\vskip-15pt
\caption{Probabilities of attachment and detachment as a function of
  optical trap well depth, at various values of trap velocity
  $v_\text{OT} = \{0.01, 0.05, 0.1, 0.5, 1 \}$. The optical trap well
  depth $\epsilon_{\text{OT}}$ was varied from 1 to 8, whilst keeping
  all other potential parameters constant ($\epsilon_{\text{M}} = 4$,
  $k_{\text{M}} = 2$ and $k_{\text{OT}} = 2$). The symbols in (a)
  and~(b) are the non-equilibrium probabilities of attachment and
  detachment $p_{\text{A}}^{\text{neq}}$ and
  $p_{\text{D}}^{\text{neq}}$, while the symbols in (c) and~(d)
  represent the equilibrium probabilities $p_{\text{A}}$ and
  $p_{\text{D}}$ obtained from non-equilibrium umbrella sampling. The
  curves in (a) to (d) are the analytical equilibrium probabilities
  $p_{\text{A}}^{\text{anal}}$ and $p_{\text{D}}^{\text{anal}}$ (as
  appropriate).}
\label{fig:umbrella}
\end{figure*}
\begin{widetext}
To formalise these definitions, it is useful to introduce the
indicator functions $\chi_{\text{A}}$ and $\chi_{\text{D}}$,
\begin{align}
\chi_{\text{A}} (x)
& = \begin{cases}
   1   & \text{if} \quad - \infty < x \le x_{\text{M}}^{\text{ub}} , \\
   0   & \text{if} \quad x_{\text{M}}^{\text{ub}}  < x <  \infty ,
\end{cases}
\\
\chi_{\text{D}} (x) =
 & = \begin{cases}
   0   & \text{if} \quad - \infty < x < x_{\text{OT}}^{\text{lb}} , \\
   1   & \text{if} \quad x_{\text{OT}}^{\text{lb}} \le x < \infty .
\end{cases}
\end{align}
The \emph{equilibrium} probabilities for the attached and the detached
states are then simply the Boltzmann averages of $\chi_{\text{A}}$ and
$\chi_{\text{D}}$, respectively. Here of course the Boltzmann
distribution corresponding to the final potential profile ($\lambda =
\lambda_{\text{f}}$) must be used:
\begin{align}
p_{\text{A}} & =  \int^{\infty}_{- \infty}  \!\! dx \, \chi_{\text{A}} (x) \, 
p_{\text{eq}}^{\lambda_{\text{f}}} (x)  
= \braket{\chi_{\text{A}}}_\text{eq}^{\lambda_\text{f}} ,
\label{probA}
\\
p_{\text{D}} &  =  \int^{\infty}_{- \infty}  \!\! dx \, \chi_{\text{D}} (x) \, 
p_{\text{eq}}^{\lambda_{\text{f}}} (x)
= \braket{\chi_{\text{D}}}_{\text{eq}}^{\lambda_{\text{f}}} .
\label{probD}
\end{align}

For the choice of potentials in the present work, it is
straightforward to determine these values analytically. Using
arguments along the lines of those in section~\ref{analfe} for the
analytical determination of free energy differences, we can show that
\begin{align}
  p_{\text{A}}^{\text{anal}} & =
  \frac{1}{Z(x_{\text{OT}} = x_{\text{OT}}^{\text{final}})}
  \, \int_{-\infty}^{x_{\text{M}}^{\text{ub}}} \! dx \, 
  \exp{\left[-\left(\dfrac{1}{2} \, k_{\text{M}} \, x^2
      -\epsilon_{\text{M}} \right)\right]} 
   = \frac{Z_{\text{A}}}{Z(x_{\text{OT}} = x_{\text{OT}}^{\text{final}})} ,
\label{probAanal} \\
  p_{\text{D}}^{\text{anal}} & =
  \frac{1}{Z(x_{\text{OT}} = x_{\text{OT}}^{\text{final}})}
  \, \int^{\infty}_{x_{\text{OT}}^{\text{lb}}} \! dx \, 
  \exp \left[-\left(\dfrac{1}{2} \, k_{\text{OT}} \, (x-x_{\text{OT}})^2 
  -\epsilon_{\text{OT}}\right)\right] 
     = \frac{Z_{\text{D}}}{Z(x_{\text{OT}} = x_{\text{OT}}^{\text{final}})} ,
\label{probDanal}
\end{align}
where the quantities $Z_{\text{A}}$ and $Z_{\text{D}}$ in the
equations above are given by
\begin{align}
\label{anal1m}
Z_{\text{A}} & =\frac{\sqrt{\pi /2}}{\sqrt{k_{\text{M}}}} 
\exp{(\epsilon_{\text{M}})} \left[\text{erf}
  \left(\frac{x_{\text{M}}^{\text{ub}} \sqrt{k_{\text{M}}}}
  {\sqrt{2}}\right) + 1 \right] ,
\\
Z_{\text{D}} & = \frac{\sqrt{\pi /2}}{\sqrt{k_{\text{OT}}}} 
\exp{(\epsilon_{\text{OT}})} \, \text{erfc}
  \left(\frac{(x_{\text{OT}}^{\text{lb}} - x_{\text{OT}}^{\text{final}})
    \sqrt{k_{\text{OT}}}}{\sqrt{2}}\right) ,
\end{align}
and $Z(x_{\text{OT}} = x_{\text{OT}}^{\text{final}})$ is given by Eqn.~(\ref{anal1}). These expressions are useful to evaluate the degree of success of the
non-equilibrium umbrella sampling technique in determining the
equilibrium probabilities $p_{\text{A}}$ and $p_{\text{D}}$ from the
non-equilibrium computer experiment. This latter analysis is done as
follows:

We denote the total number of detachment simulations with
$N_{\text{T}}$. Similarly, $N_{\text{A}}$ denotes the number of runs
where the bead ends up in the attached state ($\chi_{\text{A}} (x (t =
t_{\text{D}})) = 1$). Analogously, $N_{\text{D}}$ is the number of
runs where the bead is finally detached. If
$p_{\text{neq}}^{\lambda_{\text{f}}} \left( x(t_{\text{D}}) \right)$
is the \emph{non-equilibrium} distribution of bead positions at the
final time $t_{\text{D}}$, then the non-equilibrium probabilities of
attachment and detachment, defined by the following expressions, are
easily estimated by simulations from the ratios
$N_\text{A}/N_\text{T}$ and $N_\text{D}/N_\text{T}$, respectively:
\begin{align}
p_{\text{A}}^{\text{neq}} & =  \int^{x_{\text{M}}^{\text{ub}}}_{- \infty}  \!\! dx \, 
p_{\text{neq}}^{\lambda_{\text{f}}} (x) 
= \int^{\infty}_{- \infty}  \!\! dx \, \chi_{\text{A}} (x)
\, p_{\text{neq}}^{\lambda_{\text{f}}}  (x) 
= \braket{\chi_{\text{A}}}_{\text{neq}}^{\lambda_{\text{f}}}  
= \frac{N_{\text{A}}}{N_{\text{T}}} ,
\label{probAneq}
\\
p_{\text{D}}^{\text{neq}} & =  \int_{x_{\text{OT}}^{\text{lb}}}^{\infty}  \!\! dx \, 
p_{\text{neq}}^{\lambda_{\text{f}}} (x)
= \int^{\infty}_{- \infty}  \!\! dx \, \chi_{\text{D}} (x) \,
p_{\text{neq}}^{\lambda_{\text{f}}}  (x) 
= \braket{\chi_{\text{D}}}_{\text{neq}}^{\lambda_{\text{f}}}
= \frac{N_{\text{D}}}{N_{\text{T}}} .
\label{probDneq}
\end{align}
\end{widetext}
We now use the technique of non-equilibrium umbrella sampling to
obtain the equilibrium probabilities from the non--equilibrium
simulations. Let us outline this method in general terms:

For an observable $B$ that has been sampled by a non-equilibrium (computer) experiment, i.e., using the probability distribution $p_{\text{neq}}^{\lambda_{\text{f}}} (x)$, we simply have to multiply each data point with the ratio $p_{\text{eq}}^{\lambda_{\text{f}}} (x)/p_{\text{neq}}^{\lambda_{\text{f}}} (x)$ such that the data point is given the weight $p_{\text{eq}}^{\lambda_{\text{f}}} (x)$ rather than $p_{\text{neq}}^{\lambda_{\text{f}}} (x)$. It can be shown~\cite{Crooks:2000ke,Williams:2011ei,williams2010Non} that the ratio $p_{\text{eq}}^{\lambda_{\text{f}}} (x)/p_{\text{neq}}^{\lambda_{\text{f}}} (x)$ is nothing but the factor $ e^{- W}$, except for normalisation. Therefore, the data need to be reweighted according to the formula
\begin{equation}
\label{eq:um}
\braket{B}_{\text{eq}}^{\lambda_{\text{f}}} 
= \frac{\braket{B e^{- W}}_{\text{F}}}{\braket{e^{- W}}_{\text{F}}}.
\end{equation} 

Application of this general formula to our observables ($\chi_{\text{A}}, \chi_{\text{D}}$) yields
\begin{align}
p_{\text{A}} & = \frac{\braket{\chi_{\text{A}}  \, e^{- W}}_{\text{F}}}
{\braket{e^{- W}}_{\text{F}}} ,
\label{probAumb}
\\
p_{\text{D}} & = \frac{\braket{\chi_{\text{D}}  \, e^{- W}}_{\text{F}}}
{\braket{e^{- W}}_\text{F}} .
\label{probDumb}
\end{align} 

Simulation data generated previously for examining the influence of
well depth in Sec.~\ref{cumul} has been used here for evaluating
the usefulness of non-equilibrium umbrella sampling, for trap
velocities $v_{\text{OT}} = \{0.01, 0.05, 0.1, 0.5, 1\}$. The potential
parameters used in the simulations are as given in the caption to
Fig.~\ref{fig:cum2}, along with $x_{\text{OT}}^{\text{final}} = 6$.

The symbols in Figs.~\ref{fig:umbrella}~(a) and~(b) are the
non-equilibrium probabilities of attachment and detachment
$p_{\text{A}}^{\text{neq}}$ and $p_{\text{D}}^{\text{neq}}$,
determined from Eqns.~(\ref{probAneq}) and~(\ref{probDneq}) for
various trap velocities, while the symbols in
Figs.~\ref{fig:umbrella}~(c) and~(d) are the equilibrium probabilities
$p_{\text{A}}$ and $p_{\text{D}}$, determined by applying the umbrella
sampling procedure as expressed in Eqns.~(\ref{probAumb})
and~(\ref{probDumb}). Error bars estimated from the ten repeated
simulations are smaller than the symbol size in
Figs.~\ref{fig:umbrella}~(c) and~(d). The curves in the subfigures of
Fig.~\ref{fig:umbrella} represent the analytical equilibrium
probabilities $p_{\text{A}}^{\text{anal}}$ and
$p_{\text{D}}^{\text{anal}}$ (as appropriate), calculated from
Eqns.~(\ref{probAanal}) and~(\ref{probDanal}), respectively.

As expected, Figs.~\ref{fig:umbrella}~(a) and~(b) indicate that the
non-equilibrium probabilities are nearly identical to the equilibrium
probabilities at low trap velocities, but deviate from the latter more
and more as the trap velocity increases. Interestingly, the greatest
departure occurs for membrane and optical trap potential well depths
that are roughly equal in magnitude. Not surprisingly, the probability
of detachment is greatest for the largest optical trap well depth,
while the likelihood of remaining in the membrane potential is high at
low trap well depths. For nearly all the trap velocities, except
perhaps at $v_{\text{OT}} = 1$ (for roughly equal trap strengths),
application of umbrella sampling recovers the equilibrium probabilities
nearly perfectly.

\section{Conclusions}
\label{sec:conclusion}

A simple model for the detachment of a ligand coated bead with the
help of an optical tweezer, from receptors on the surface of a cell to
which it is bound, has been used to examine if fluctuation theorems
are useful in determining equilibrium free energies, which in turn
provide information about the binding energetics. By using truncated
harmonic potentials to represent the stationary cell membrane and the
moving optical trap, and a Langevin equation to model the stochastic
motion of the bead in these potentials, the distribution of work
performed in driving the system from an initial equilibrium state to a
final non-equilibrium state (at various finite rates) has been
calculated by carrying out repeated simulations of the Langevin
equation in the forward and reverse directions. The former corresponds
to the membrane and trap potentials being superposed at time $t=0$,
followed by the optical trap being translated uniformly until the two
potentials are sufficiently apart at the final time
$t=t_{\text{D}}$. The latter refers to the opposite situation.

The calculation of work distributions enables the determination of the
equilibrium free energy change between the initial and final states of
the system, using both the Crooks fluctuation theorem and the
Jarzynsky equality. The simplicity of the model also permits a
straight forward determination of the exact free energy change by
analytical means. It is found that both fluctuation theorems lead to
excellent predictions provided the rate of switching from the initial
to the final state is sufficiently slow. For relatively rapid rates of
trap translation, sampling problems (for the given sample size) lead
to a decrease in accuracy. The reduction in accuracy is discussed both
in terms of a Gaussian approximation for the work distributions, and a
cumulant expansion for the average of the exponential of work.

The method of non-equilibrium umbrella sampling has been used to
determine the equilibrium probability that, after translating the trap
from its initial to its final location, the bead and cell are still
attached (i.e., the bead lies only within the range of influence of the
membrane potential), and the equilibrium probability that the bead and
cell have been detached (the bead lies only within the range of
influence of the optical trap potential), for a range of different
values of the optical trap well depth. It is seen that by
appropriately analysing the non-equilibrium simulation data, accurate
estimates of the equilibrium probabilities of attachment and
detachment can be found for all but the highest rates of trap
translation.

In conclusion, although a very simple model has been used, the present work demonstrates that non-equilibrium fluctuation theorems can be applied without significant statistical problems to binding/unbinding experiments carried out with optical trap velocities that are realizable under experimental conditions. Combined with the theoretical procedure outlined in section~\ref{fluct}, they could provide a reliable means of extracting unknown membrane potentials (see Eqn.~(\ref{convol})).

\section{Acknowledgements}
\label{sec:Ak}
The authors gratefully acknowledge CPU time grants from the National
Computational Infrastructure (NCI) facility hosted by the Australian
National University, and Victorian Life Sciences Computation
Initiative (VLSCI) hosted by the University of Melbourne. We thank
C. Sasmal for help with preparation of some of the figures.

\bibliography{EmmaBib} 
\bibliographystyle{rsc} 

\end{document}